\documentclass[a4paper,fleqn,usenatbib]{mnras}

\usepackage[T1]{fontenc}
\usepackage{ae,aecompl}

\usepackage{graphicx}	
\usepackage{amsmath}	
\usepackage{amssymb}	

\title[Young population in the galaxy NGC~253]{Identification and analysis of the young population in the starburst galaxy NGC~253}
\author[M. J. Rodr\'\i{}guez et al.]{
M. J. Rodr\'\i{}guez,$^{1}$\thanks{E-mail: jimenaro@fcaglp.unlp.edu.ar}
G. Baume$^{1,2}$ and 
C. Feinstein $^{1,2}$
\\
$^{1}$Instituto de Astrof\'\i{}sica de La Plata (CONICET-UNLP), Paseo del bosque S/N, 
La Plata (B1900FWA), Argentina\\
$^{2}$Facultad de Ciencias Astron\'omicas y Geof\'\i{}sicas - Universidad Nacional de La Plata, Paseo del bosque S/N, 
La Plata (B1900FWA), Argentina\\
}
\date{Accepted XXX. Received YYY; in original form ZZZ}

\pubyear{2017}

\hypersetup{draft}  
\begin{document}
\label{firstpage}
\pagerange{\pageref{firstpage}--\pageref{lastpage}}
\maketitle
\begin{abstract}
We present a study of the young population in the starburst galaxy NGC~253. In particular, we focused our attention on searching young star groups, obtaining their main properties and studying their hierarchical organization. For this task, we used multiband images and their corresponding photometric data obtained with the Advanced Camera for Surveys of the Hubble Space Telescope (ACS/HST). 

We have first derived the absorption affecting the different regions of the galaxy. Then, we applied an automatic and objective searching method over the corrected data in order to detect young star groups. We complemented this result with the construction of the stellar density map for the blue young population. A statistical procedure to decontaminate the photometric diagrams from field stars was applied over the detected groups and we estimated their fundamental parameters.

As a result, we built a catalog of 875 new identified young groups with their main characteristics, including coordinates, sizes, estimated number of members, stellar densities, luminosity function (LF) slopes and galactocentric distances. We observed these groups delineate different structures of the galaxy, and they are the last step in the hierarchical way in which the young population is organized. From their size distribution, we found they have typical radius of $\sim 40-50$ pc. These values are consistent with those ones found in others nearby galaxies. We estimated a mean value of the LF slope of 0.21 and an average density of 0.0006 stars/pc$^3$ for the identified young groups taking into account stars earlier than B6.
\end{abstract}


\section{Introduction}
\label{intro}

Young star groups exist in a wide range of sizes, from compact star clusters to star complexes going through OB associations. Young open clusters contain a few tents to $10^{5}$ stars in a typical diameter of a few parsecs and they are gravitationally bound \citep{2016arXiv160700027M}. OB associations are composed of young massive stars formed from the same molecular cloud, they are gravitationally unbound with low densities.

The distribution and properties of these systems are useful to understand the most recent history of the host galaxy as well as the star formation process under different environments. For this reason, in the last decades several works has been focused in performing global studies and assembling catalogs of young stars groups over the nearest galaxies: 
LMC \citep{2003A&A...405..111G}, 
SMC \citep{1991A&A...244...69B}, 
M~31 \citep{2012AJ....144..142B}, 
M~33 \citep{1999ApJS..122..431C, 2005A&A...444..831B}, 
NGC~6347 \citep{2015A&A...573A..95M}, 
NGC~300 \citep{2001A&A...371..497P,2016A&A...594A..34R}, 
NGC~7793 \citep{2005A&A...440..783P}, 
M~81 \citep{2010AJ....139.1413N}, 
M~101 \citep{1996AJ....112.1009B}, 
to name just a few. 

In particular, the Sculptor Group galaxy NGC~253, is a barred almost edge-on spiral galaxy \citep[SAB(s)c D,][]{2015MNRAS.446..943V}, at a distance of 3.56 Mpc \citep{2013AJ....146...86T}, which correspond to a projected linear scale of $\sim$~17~pc/arcsec. One of its most outstanding features is the starburst activity. Different works indicate that these bursts are located in the galactic central region up to $\sim$~300~pc in radius, with a star formation rate (SFR) of $\sim$~2~-~3~M$_{\sun}$~yr$^{-1}$\citep[]{2001A&A...377...73R, 2005ApJ...629..767O}, however their nature is not yet well understood. According to \cite{1998ApJ...505..639E} the bursts are caused by the presence of the bar, that drives the gas into the central region. Additionally, \cite{2010ApJ...725.1342D} suggested that the interaction with a now defunct companion occurred  within the past $\sim$0.2 Gyr, stimulating the formation of the bar and starburst activity.

Several young clusters were identified in the nuclear star forming region of the galaxy \citep[e.g.][]{1996AJ....112..534W,2009MNRAS.392L..16F}. One of them was associated with a super star cluster of $\sim$1.4$\ x\ 10^{7}~M\odot$ and an estimated age of 5.7~Myr. The simultaneous presence of red supergiants and OB stars in this cluster reveals several epochs of star formation \citep{2009ApJ...697.1180K}. \cite{2016ApJ...818..142D} found distinct substructures in it, suggesting that it is a star forming complex and not a single cluster.

The starburst activity added to its proximity make this galaxy an interesting candidate to study its young star population. In spite of this, there has not been made a detection of young star structures outside its central region. In this work, we carried out for first time the search and identification of young groups in NGC~253 adopting an automatic and objective searching method, and its subsequent analysis. We took advantage of excellent quality data from the Hubble Space Telescope (HST) that cover approximately the north-east side and the center of the galaxy (see Fig. \ref{campos}).

The paper is organized as follows: In Sect.~\ref{data} we describe the observations and data sets, together with the reduction and data set-up. In Sect.~\ref{search} we presented the methods we used to search for and identify young stellar structures. Section~\ref{analysis} presents our analysis of the detected groups. We discuss our results in Sect.~\ref{discussion}. Finally, in Sect.~\ref{conclusions} we summarized our work and remark our main results.

\section{Data}
\label{data}

\subsection{Observations}

\begin{figure}
\resizebox{\hsize}{!}{\includegraphics{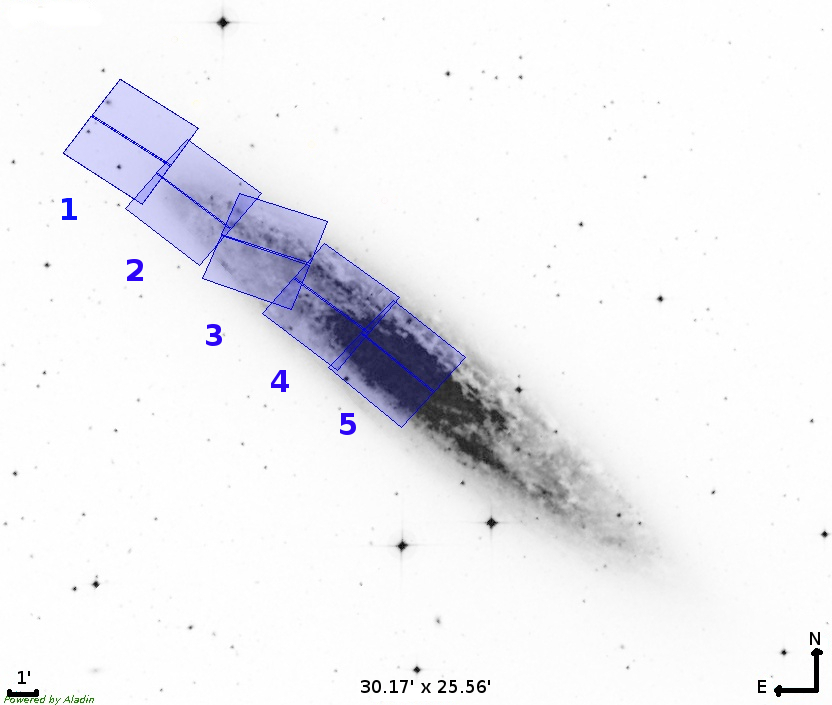}}
\caption{Distribution of the different HST/ACS fields (rectangles) used in this work overlaid on a Digitized Sky Survey (DSS) image of NGC~253. The North is up and East is left, the image cover $30\farcm1 \times 25\farcm5$.}
\label{campos}
\end{figure}

The images used in this work were acquired from the Hubble Legacy Archive \footnote{http://hla.stsci.edu/} and they were obtained in September 2006, as part of the program GO-10915 (PI: J. Dalcanton), during HST Cycle 15. These observations correspond to five fields of NGC~253, and were carried out with the Wide Field Camera (WFC) of the Advanced Camera for Surveys (ACS). The WFC has a mosaic of two CCD detectors with a field of view of $3\farcm3~\times~3\farcm3$ and a scale of $0\farcs049$/pixel. The five fields cover a total length of approximately 16' encompassing more than half of the galaxy (see Fig.~\ref{campos}). Three broadband filters were used ($F475W, F606W$ and $F814W$), with total exposure times of 1482~s., 1508~s. and 1534~s. respectively, excepted the Field~1, which has higher exposure times: 2256~s.($F475W$), 2283~s. ($F606W$) and 2253~s. ($F814W$).

\subsection{Photometry}

\begin{figure}
\resizebox{\hsize}{!}{\includegraphics{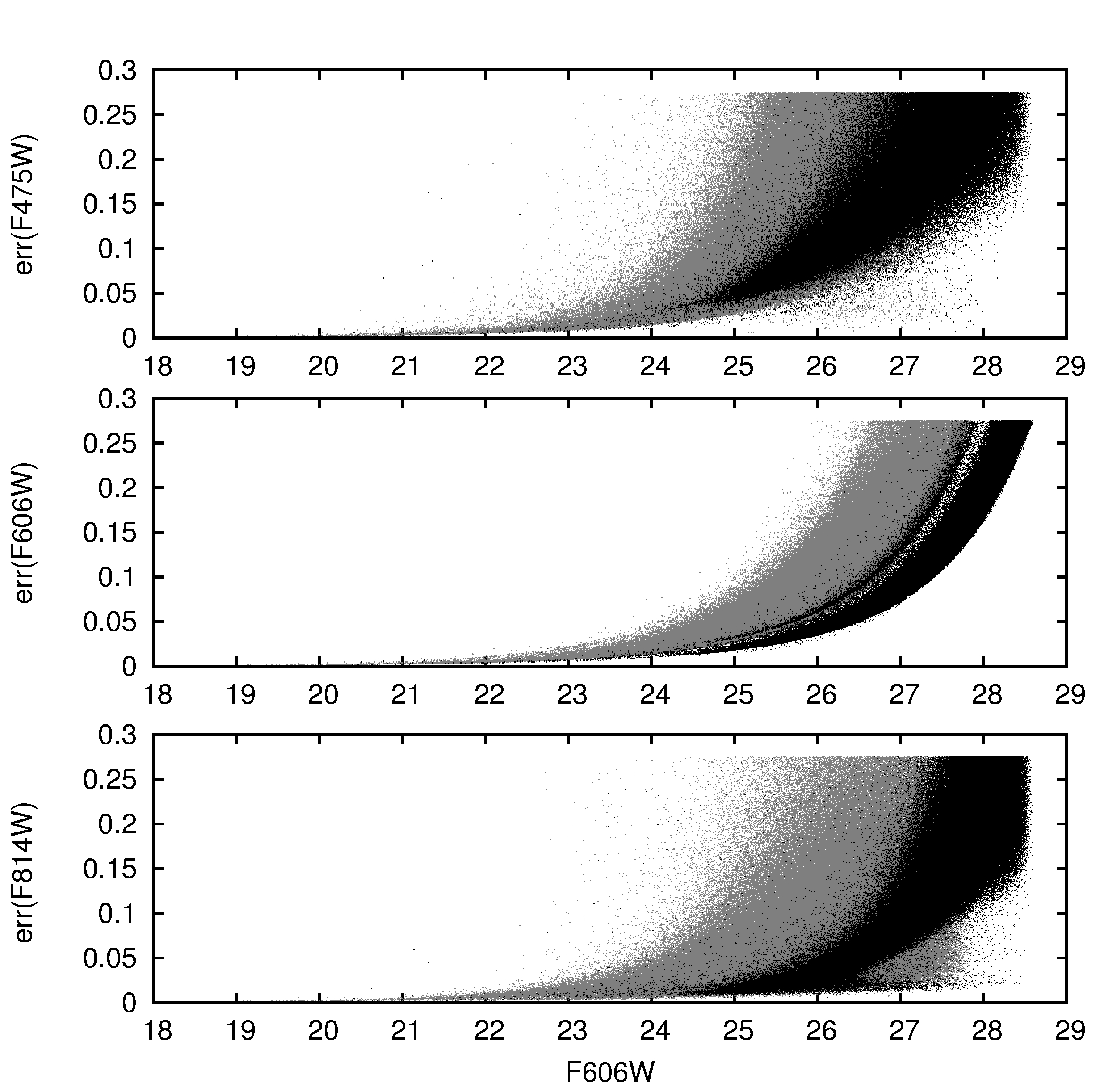}}
\caption{Photometric errors in the bands $F475W$ (top), $F606W$ (middle) and $F814W$ (bottom) vs. $F606W$ magnitude for the fields 1 (black) and 5 (grey).}
\label{errors}
\end{figure}

\begin{figure}
\resizebox{\hsize}{!}{\includegraphics{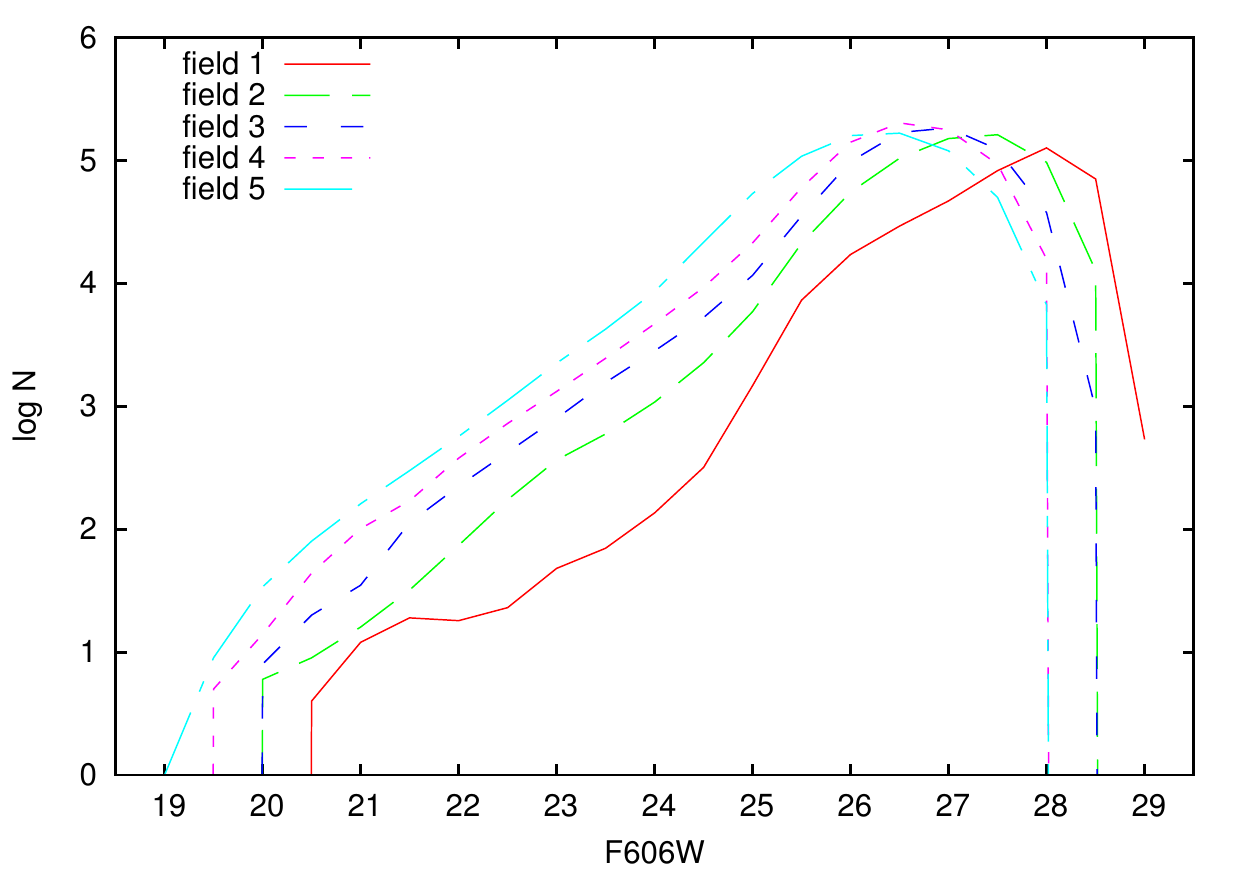}}
\caption{Observed LFs. There is a curve for each studied field.}
\label{LF-total}
\end{figure}

Binary FITS tables of photometry were obtained from the database of the Space Telescope Science Institute (STScI)~\footnote{MAST: https://archive.stsci.edu/}. They correspond to the ``star files" from the ACS Nearby Galaxy Survey (ANGST). These files contain the photometry of all objects classified as stars with good signal-to-noise values ($S/N~>~4$) and data flag~$<$~8. These data were obtained performing point spread function (PSF) photometry using the package DOLPHOT adapted for the ACS camera. The reduction procedure was explained in detail in \citet{2008glv..book..115D}. In Fig.~\ref{errors} we show the photometric errors in the different bands for the field 1 and 5, which are, respectively, the fields with the lowest and highest stellar density. To evaluate the completeness of the data, we built the luminosity function (LF) for the five studied fields (see Fig.~\ref{LF-total}). The number of stars per bin starts to decrease at $F606W~\sim~26.5$. Therefore we considered that the sample is complete up to this value. 

\subsection{Photometric correlation of tables}
\label{catalog}

Three photometric tables per field (in total 15 tables) were obtained from the ANGST, each one provide photometric information of only two bands. In order to join the three magnitudes in a single table for field, we used the code STILTS\footnote{http://www.star.bris.ac.uk/$\sim$mbt/stilts/} to perform a cross-correlation (with logical "OR") between tables with $F606W-F475W$ bands and those with $F814W-F606W$ bands, obtaining five photometric tables, one for each field.

Furthermore, the adjacent fields slightly overlap each other (see Fig.~\ref{campos}), therefore we used STILTS to join the information of the five fields in one single table with approximately 3~x~$10^6$ objects. For this purposes, for the objects in the overlap region we took a final magnitude that results from the average between the stars in each field. Then we slightly shifted the coordinates of all the field to agree with those given at Gaia Data Release 1 (\citet{2016A&A...595A...2G}; see Table~\ref{shift}). 

\begin{table}
\caption{Astrometric (top table) and photometric (bottom table) corrections applied 
to obtain the final catalog (see text for details).}             
\label{shift}      
\centering          
\begin{tabular}{c c c }     

\hline\hline       
                      
Field & $\Delta\alpha \cos(\delta) ["]$ & $\Delta\delta ["] $\\ 
\hline 
  5-gaia & -0.056  &   0.078 \\     
   4-5   &  0.278  &  -0.480 \\
   3-4   &  0.252  &   0.285 \\
   2-3   & -0.022  &   0.022 \\
   1-2   & -1.055  &  -0.096 \\
\hline       
\multicolumn{3}{l}{The first column is: (field shifted)-(field to coincide).}\\
\end{tabular}

\begin{tabular}{c c c c c}

\hline\hline 
Field &    $\Delta F475W$    &       $\Delta F606W$ &  $\Delta F814W$ & N\\
\hline
   4-5  &  0.011 $\pm$  0.082  & 0.036 $\pm$  0.073  &  0.047 $\pm$  0.081  &  221 \\
   3-4  &  0.119 $\pm$  0.183  & 0.081 $\pm$  0.115  &  0.047 $\pm$  0.099  &   45 \\
   2-3  & -0.076 $\pm$  0.119  &-0.015 $\pm$  0.082  & -0.003 $\pm$  0.067  &  162 \\
   1-2  & -0.001 $\pm$  0.015  &-0.042 $\pm$  0.017  &  0.047 $\pm$  0.017  &    2 \\
\hline 
\multicolumn{5}{l}{N is the number of stars in the two fields.}
\\
\end{tabular}
\end{table}

\section{Identification of young groups}
\label{search}

As a previous step to identify star groups over the galaxy field, it was necessary to study a couple of problems present in our data sample. They were the following ones:

\begin{enumerate} 
\item Notwithstanding, the ACS/HST provide images with a high spatial resolution ($\sim~0\farcs05$), it is possible that, in the crowded regions of NGC~253, some objects identified as single sources are really blends of two or more ones, or even compact star clusters.
\item NGC~253 is heavily obscured at visible wavelengths, specially in its central region, due to its large amounts of dust \citep{1980ApJ...239...54P}. This causes that the detected objects appear redder than its real color, making it difficult to identify young blue star populations.
\end{enumerate}

\subsection{The blending problem}
\label{blending}

Several previous works have studied this problem (e.g. \citealt{1998AJ....115.2459R}; \citealt{2005MNRAS.358..883K}). In our case, to obtain a measure of the importance of the blending effect, we followed the reasoning presented by \cite{2005MNRAS.358..883K}. Therefore, we performed several numerical simulations using random distributed stars with a uniform spatial distribution, a power law distribution in bright, ranging from 18 to 27, and a uniform distribution in color, ranging from 0 to 2. We analyzed different stellar densities cases ranging from 1~star/arcsec$^2$ to 10~stars/arcsec$^2$. This range of values correspond to those found along the galaxy, but remarking that the highest ones only were present in a few special places of the galaxy (see in advance Fig.~\ref{Dmap}). Therefore, for a given stellar density value, we build the corresponding bidimensional spatial histograms with a binning step of $0\farcs07$ to obtain binned areas that simulate the ACS/HST resolving power. We could then evaluate the total amount of objects with blending in each simulation. To evaluate the error in magnitudes and colors due to this effect, we computed the total magnitude and color on each bin and we compared these values with the corresponding ones to the brightest star in that bin. The above procedure was repeated five hundred times for each stellar density. As a result, we obtained that the error in magnitude, as expected, have an increasing behavior with the stellar density value. Therefore, in the worst case (10~stars/arcsec$^2$), we found that the percentage of stars that had an error in magnitude greater than 0.1~mag increased from almost null values for bright stars ($F606W \sim 18$) to only 2\% for stars of magnitude $F606W = 24$. Regarding the behavior in color, we found, again for 10~stars/arcsec$^2$, that at most 15\% of the stars have errors in colors larger than 0.1~mag and this value was approximately independent of the magnitude value. These obtained figures for both errors in magnitude and colors revealed that notwithstanding the blending effect was still present using ACS/HST data, the amount of stars that reached errors of 0.1~mag were only a small fraction of all the objects, at least for the studied stellar densities.

\subsection{The absorption problem}
\label{av}

To estimate the visual absorption ($A_V$) along the galaxy, we selected the brightest objects ($F606W < 24$) and compared their locations on the color-magnitude diagrams (CMDs) with the theoretical evolutionary model PARSEC version 1.2S \citep{2014MNRAS.445.4287T} with Z=0.0152 corresponding to 10$^{7}$yr. Assuming as a first approximation that all the selected brightest objects belong to the main sequence, it was possible to deredden their location on the CMDs and estimate approximately their individual absorption values. This procedure could include intrinsically red stars that would produce overestimated $A_V$ values. Therefore, we minimized this effect, and a possible blending effect, assigning to each star the minimum $A_{V}$ value in a surrounding spatial region with at least 10 stars. We adopted a single isochrone to obtain the absorption values for a sample of stars that could cover a wide range of ages and/or metallicities.

The results of the indicated procedure could be sensible to the crowding/blending present in the galaxy and/or to the adopted isochrone model. To check the former issue we computed the absorption maps for three $F606W$ magnitude ranges (8.8 - 23.2; 23.2 - 23.7 and 23.7 - 24.0) in such way that we have the same amount of stars (11222) for each group. The obtained absorption maps did not show significant differences among themselves or with the map obtained considering the entire range of bright stars ($F606W < 24$). The comparison of the different maps allowed us to estimate an error in our results $e_{A_V} \sim 0.09$ mag. On the other hand, our numerical simulations (see Sect.~\ref{blending}) and the use of different isochrones revealed that the employed procedure to estimate the absorption values along the galaxy provides enough precision for a first search of young star groups.

\begin{figure*}
\resizebox{\hsize}{!}{\includegraphics[]{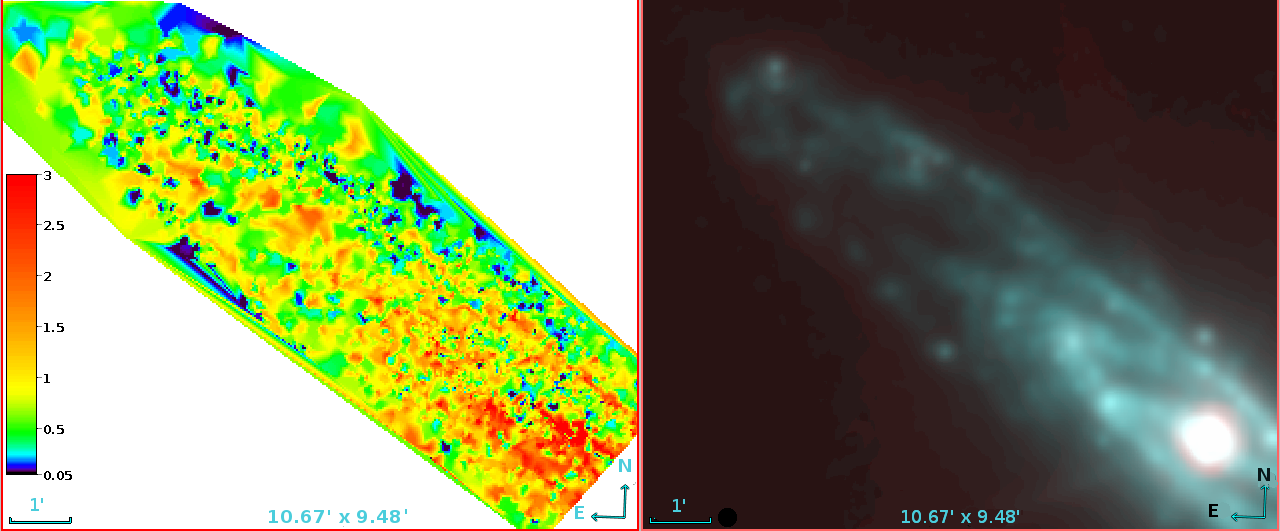}}
\caption{\textit{Left:}Visual absorption ($A_V$) map of the observed region of NGC~253. \textit{Right:} WISE IR color image of NGC~253, obtained by the combination of the W3 and W4 bands. A color version of this figure and a FITS version of the absorption map are available online.}
\label{Av}
\end{figure*}

In the left panel of Fig.~\ref{Av} we present the obtained absorption map of the observed region of NGC~253 following the above procedure. In this figure we did not include the most external field (Field 1, see Fig \ref{campos}), because it presented several foreground stars or background galaxies and very few bright stars belonging to the galaxy, so it introduce many spurious detections. The obtained absorption map was used then to correct the observed magnitudes of the brightest objects. This is: $(F475W)_0 = F475W - A_{F475W}$; $(F606W)_0 = F606W - A_{F606W}$ and $(F814W)_0 = F814W - A_{F814W}$. For this task, we use the corresponding absorption value for each filter, that were estimated using the coefficients $A_{F475W}/A_V$~=~1.192, $A_{F606W}/A_V$=~0.923 and $A_{F814W}/A_V$=~0.605\footnote{http://stev.oapd.inaf.it/cgi-bin/cmd} \citep{1994ApJ...437..262O}. In the right panel of Fig.~\ref{Av} we show a WISE IR false color image of the same NGC~253 region shown in the left panel. It was obtained from the combination of the W3 and W4 bands, which best map the dust content. We could note that the regions with stronger absorption correspond to the brightest areas (highest dust content), they are the nucleus, the bar, and the most inner spiral arms.

\subsection{Selecting blue and red objects}
\label{blue}

\begin{figure}
{\includegraphics[scale=0.3]{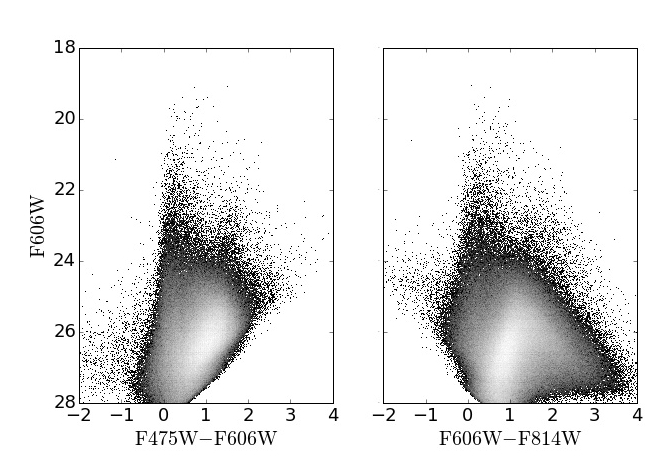}}
{\includegraphics[scale=0.3]{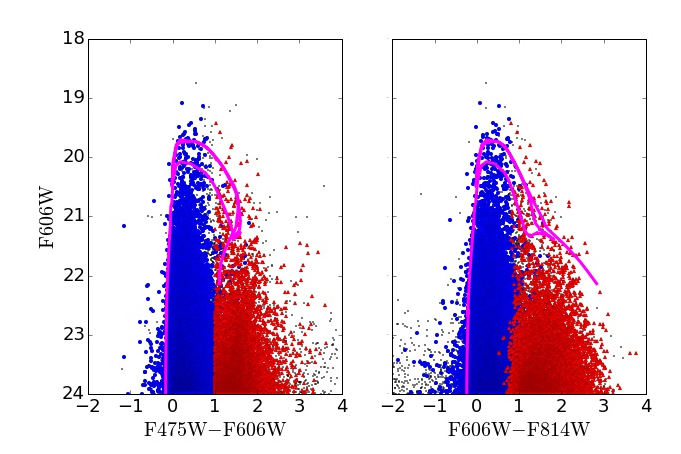}}
\caption{{\it Top panel: } Combined Hess diagrams and CMDs with all the detected objects over the total covered region in NGC~253. The different grey shades indicates the stars relative density from black (less dense) to white (more dense)
{\it Bottom panel:}
Blue circles and red triangles indicate the selected brightest blue and red objects (see Sect.~\ref{blue}). The pink line indicate the PARSEC version 1.2S isochrone \citet{2014MNRAS.445.4287T} corresponding to 10$^7$ yr and metallicity Z=0.0152, displaced adopting a distance modulus of $(Vo-M_{V})=27.75$ \citep{2013AJ....146...86T}, a normal reddening law ($R = A_{V}/E(B-V) = 3.1$) and a value for $E(B-V) = 0.016$ corresponding to the foreground reddening toward NGC~253 \citep{2011ApJ...737..103S}. A color version of this figure is available online.}
\label{cmd-t}
\end{figure}

To split the selected brightest objects between intrinsic blue ones and those probably identified with red stars, we used the following criteria: a) $F475W_0 - F606W_0 < 0.5$ and $F606W_0 - F814W_0 < 0.5$ for adopted $blue~objects$, and b) $F475W - F606W > 1$ and $F606W -F814W > 1$ for adopted $red~objects$. For this last group we considered only those objects that were not include in the blue sample. It is important to note that the first group were selected over the corrected magnitudes, the red objects instead were selected using the observed magnitudes. We used then the $blue~objects$ to detect young star groups. A combination of Hess diagrams and CMDs diagrams of all the detected objects are showed in the top panel of Fig.~\ref{cmd-t}. In the bottom panel we showed the the CMDs of each selected group where the diffuse border between $blue$ and $red~objects$ reveals the strong dispersion of the absorption values present along the galaxy.

\subsection{Searching method}
\label{secPLC}

\begin{figure}
\resizebox{\hsize}{!}{\includegraphics{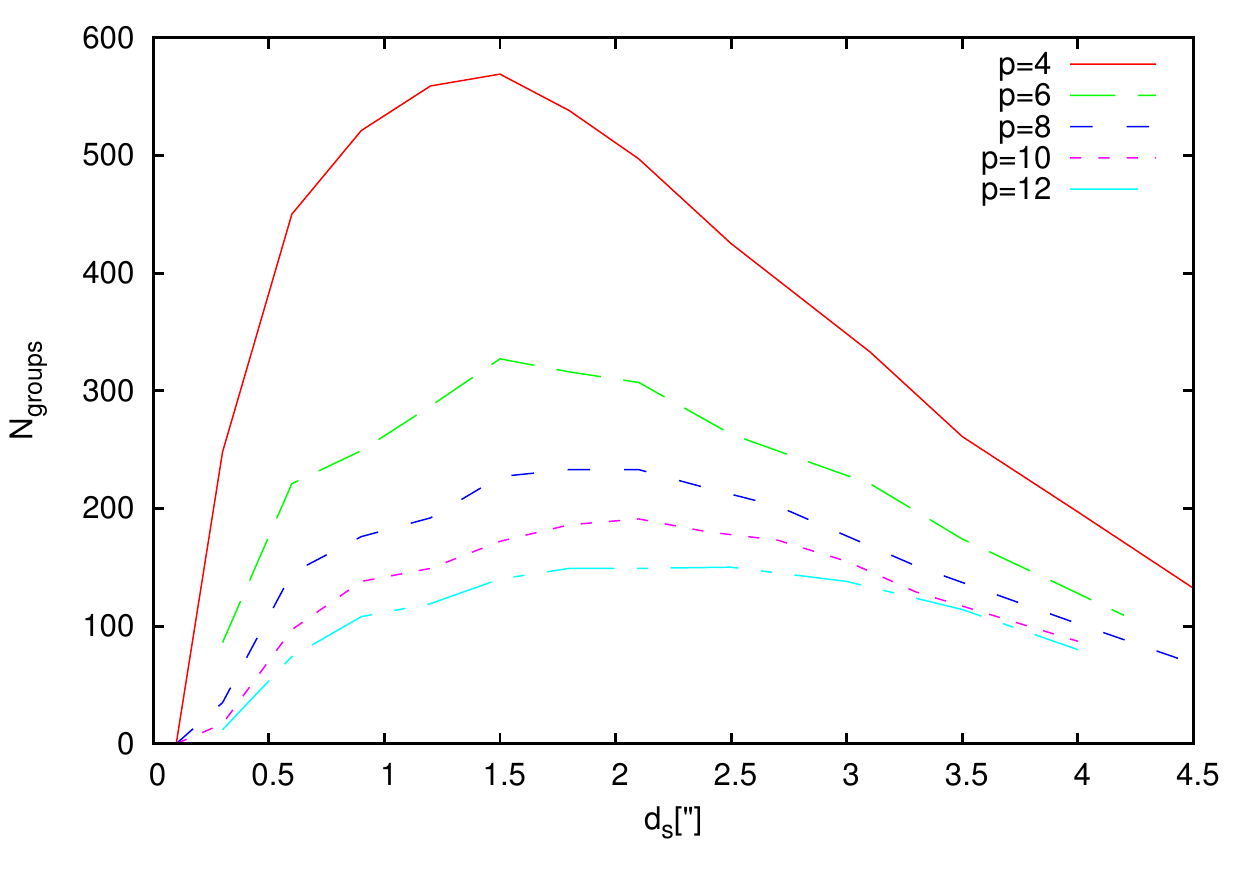}}
\caption{Behavior of the number of groups detected using the PLC technique as a function of the parameter $d_{s}$ for different values of the parameter $p$.}
\label{plc}
\end{figure}

With the aim of identify young star groups in NGC~253, we employed the Path Linkage Criterion \citep[PLC][]{{1991A&A...244...69B}} over the \textit{blue objects} selected above. The main idea of the technique is that two objects belong to the same star group if it is possible to connect them by successive links of $blue~objects$. The link distance between selected objects has not to be larger than a fixed parameter called search radius $d_{s}$. A young star group is detected when it is possible to link more than $p$ objects. To adopt adequate values for the parameters $d_s$ and $p$, we plot in Fig.~\ref{plc} the number of identified groups using the PLC as a function of $d_s$ for different $p$ values. A good choice for the parameter $d_s$ is set by the maximum number of groups detected for a given $p$ value \citep{2001A&A...371..497P}. Figure \ref{plc} reveals this maximum is located for $d_s$ between 1.5-2~arcsec. Based on these values, and with the caution of detect the smallest subgroups of each large association or stellar complex, we adopted a more extended range of $d_s$, covering the range 0.3-2~arcsec \citep[$\sim$ 5-34 pc at 3.56 Mpc;][]{2013AJ....146...86T}. Using this criteria, the method identified first the smallest groups using $d_s = 0.3$~arcsec, then this value was increased using a step of 0.4~arcsec and the PLC method was run again over the remaining \textit{blue objects}. This procedure was repeated until $d_s$ reached 2~arcsec value. Regarding the parameter $p$, it must be chosen prudently. While a low value results in many spurious detections, a high one may cause the lost of the smallest groups. We studied then the obtained results with different $p$ values (see Fig. \ref{plc}). We also noticed that in a similar study for NGC~300 galaxy, \citet{2016A&A...594A..34R} adopted $p = 10$ stars, however for NGC~253 we are leading with a more distant galaxy and there is  probably a blending effect of the individual objects in the used ACS images. Therefore, we considered $p = 8$ objects as a reasonable value.
 
\begin{figure}
\resizebox{\hsize}{!}{\includegraphics{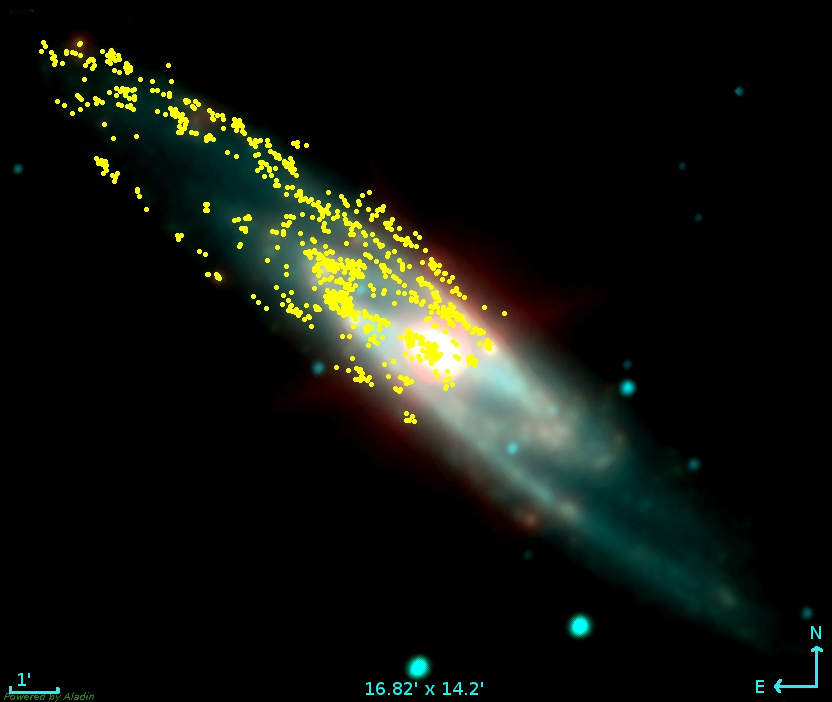}}
\caption{Detected young groups (yellow symbols) over an infrared WISE false color image of NGC~253. The image cover $16\farcm82 \times 14\farcm2$ the North is up and East is left. A color version of this figure is available online.}
\label{asoc}
\end{figure}

Employing the described procedure, we detected 897 individual groups of young objects. However, a few of these groups were spurious detection caused bright foreground stars, which caused noise spikes in the extended PSFs, or background galaxies, which were identified as faint extended diffuse objects. These false detections were located mostly in Field 1 and in Field 5. Removing these spurious detections manually, we obtained a final list of 875 young star groups. These groups are show in Fig.~\ref{asoc} over an infrared WISE\footnote{http://wise.ssl.berkeley.edu/} false color image of the galaxy. It can be noted the good agreement between their distribution and large galactic structures as the galactic spiral arms and the central bar (see Sects.~\ref{densitymap} and \ref{distribution}).

\subsection{Stochastic detections}

To evaluate the fraction of star groups found using the PLC that could be stochastic gatherings of stars and not real groups, we ran numerical simulations using randomly distributed sources with similar star densities to three selected regions, which have very different stellar density values. These are: 1) the region at $\sim$2 arcmin NE of the galactic nucleus, 2) a place in one of the spiral arms and 3) a region between the spiral arms. In each region we measured the mean stellar density of the \textit{blue objects}, and we respectively found the following values:
1) $\rho$~= 0.635~stars/arcsec$^2$,
2) $\rho$~= 0.203~stars/arcsec$^2$ and
3) $\rho$~= 0.028~stars/arcsec$^2$.

The first region is the densest, close to 0.64 \textit{blue objects} per arcsec. This high density means that the average distance among stars is  0\farcs70. So, running PLC algorithm to similar search radius or greater the method will find all the star belonging to one single cluster that is a collection of all the stars. Using 0\farcs3, the lowest search radius that we have used for the real data, we found one stochastic cluster in four cases from 10000 experiments, a very low number. Increasing the $d_s$ value, the obtained numbers were still low, for example for $d_s = 0\farcs4$ we found 149 clusters in 10000 experiments. We noticed that the stellar groups found by PLC over the numerical simulations had '; ``worm" shapes. This means that they seemed as extended twisted line, and they did not have any circular shape as would be expected for real star groups. Therefore, most of the stochastic clusters did not pass a simple visual inspection. We did not found any stochastic cluster for the cases 2) and 3) for $d_s = 0\farcs3$ and a very few for higher search radius.

\subsection{Identifying large young structures}
\label{densitymap}

Larger young stellar structures could be found using the PLC with even larger searching radius or building the stellar density map. We choose this last option since it produces better and clear results.

To build the corresponding density map we constructed a two dimensional histogram counting the numbers of objects in spatial bin sizes $ 8 \times 8$~arcsec$^2$. Then, we applied the drizzle method considering a 2.0 arcsec step \citep[see][for details of this method]{2002PASP..114..144F}. Over this map we plotted isodensity at several values: 40, 80, 110, and 145 stars per bin of $8 \times 8$~arcsec$^{2}$. 

\begin{figure}
\resizebox{\hsize}{!}{\includegraphics{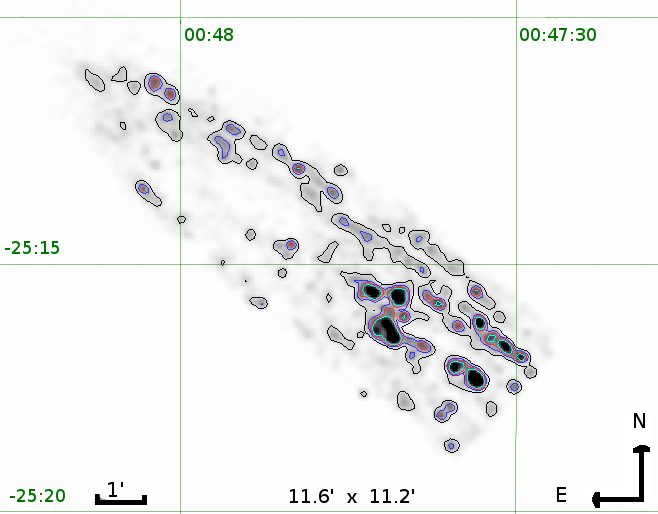}}
\caption{Stellar density map of blue stars, the overlapping contours correspond to different density levels: 40, 80, 110, 145 stars per bin of $8~\times~8$~arcsec$^{2}$. The image field is $11\farcm6~\times~11\farcm2$, the North is up and East is left. A color and a FITS versions of this figure are available online.}
\label{Dmap}
\end{figure}

The stellar density map with the different contours is presented in Fig.~\ref{Dmap}. Here we also can identify how the blue population delineates the spiral arms of the galaxy. Furthermore, the way in which larger structures enclose smaller and denser ones point out a hierarchical behavior of the young population. We would discuss this result in Sect.~ref{hierarchical}. 

\section{Analysis}
\label{analysis}

We performed an automatic analysis of the detected young star groups using a numerical code developed in FORTRAN 95. This code allowed the study of selected groups in a systematic and homogeneous way. The code estimated the basic parameters of each group detected by the PLC, as their sizes, densities, amount of objects members, and build their CMDs and LFs, with their corresponding slope values.  

Table~\ref{tabla1} presents the first ten rows of the resulting catalog, in which are listed the properties for each of the 875 young groups. The complete version is only available online.

\subsection{Sizes and densities}

The location and size of each group ($\alpha_{J2000}$, $\delta_{J2000}$ and radius, $r$) were computed, respectively, as the mean and two times the radial standard deviation ($\sigma$) of the location of the corresponding objects identified by the PLC method. An approximate value of the groups stellar density could be estimated counting the total numbers of stars in a volume of $\pi~r^{3}$ (see Table~\ref{tabla1}).




\subsection{Field star decontamination}

To better study each group, our code decontaminated their corresponding CMDs form field stars preforming a statistical cleaning. The clean method was based in the comparison between the CMD for a young group region and the CMD for a corresponding field region located near that group and covering the same sky area. Therefore, those objects with similar positions over both CMDs were eliminated in the CMD of the group region \citep[see][for details]{2003AJ....125..742G}. 

The above procedure was applied taking into account the color indexes $CI1 = F475W-F606W$ and $CI2 = F606W-F814W$ simultaneously. For each star in the region we calculated the distance in the CMDs with all the star in the field. The most similar star in the field provided the minimum distance value, if this value was less than a given tolerance, then the star in the group region was subtracted. The CMD distance was calculated using the following expression:

\noindent$\{(F606W_{r}-F606W_{f})A\}^2 +\\ \{(CI1_{r}-CI1_{f})B\}^2+\{(CI2_{r}-CI2_{f})B\}^2$

\noindent where the subscripts $``r"$ and $``f"$ refers to stars in the region and field respectively and the constants $A$ and $B$ are normalization factors.

In order to obtain a field region as homogeneous as possible, avoiding possible neighbor groups or to fall outside the field of view. The code choose five different field regions for each group. The first region is a ring around the studied group, and the others are selected as follows: ($\alpha_{0}\pm\Delta\alpha $, $\delta_{0}$); ($\alpha_{0}$, $\delta_{0}\pm\Delta\delta$) where $\Delta\alpha=\Delta\delta=11\farcs6\sim~200$~pc. Then the fields with the maximum and minimum number of stars were discarded, and the final decontamination result is an average of the decontamination obtained with the three remaining field regions.


\subsection{Color-magnitude diagrams}
\label{cmd_sect}

\begin{figure}
{\includegraphics[scale=0.65]{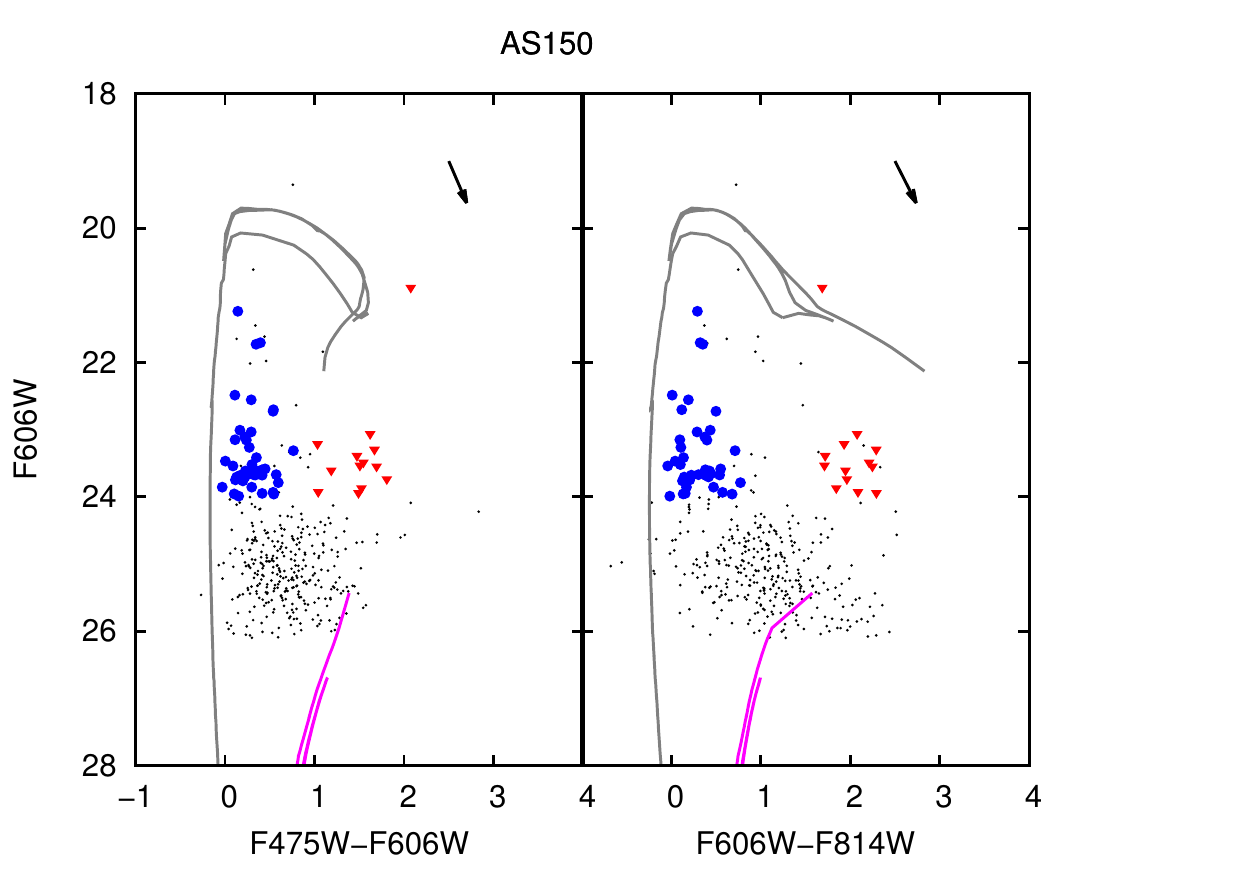}} \\
{\includegraphics[scale=0.65]{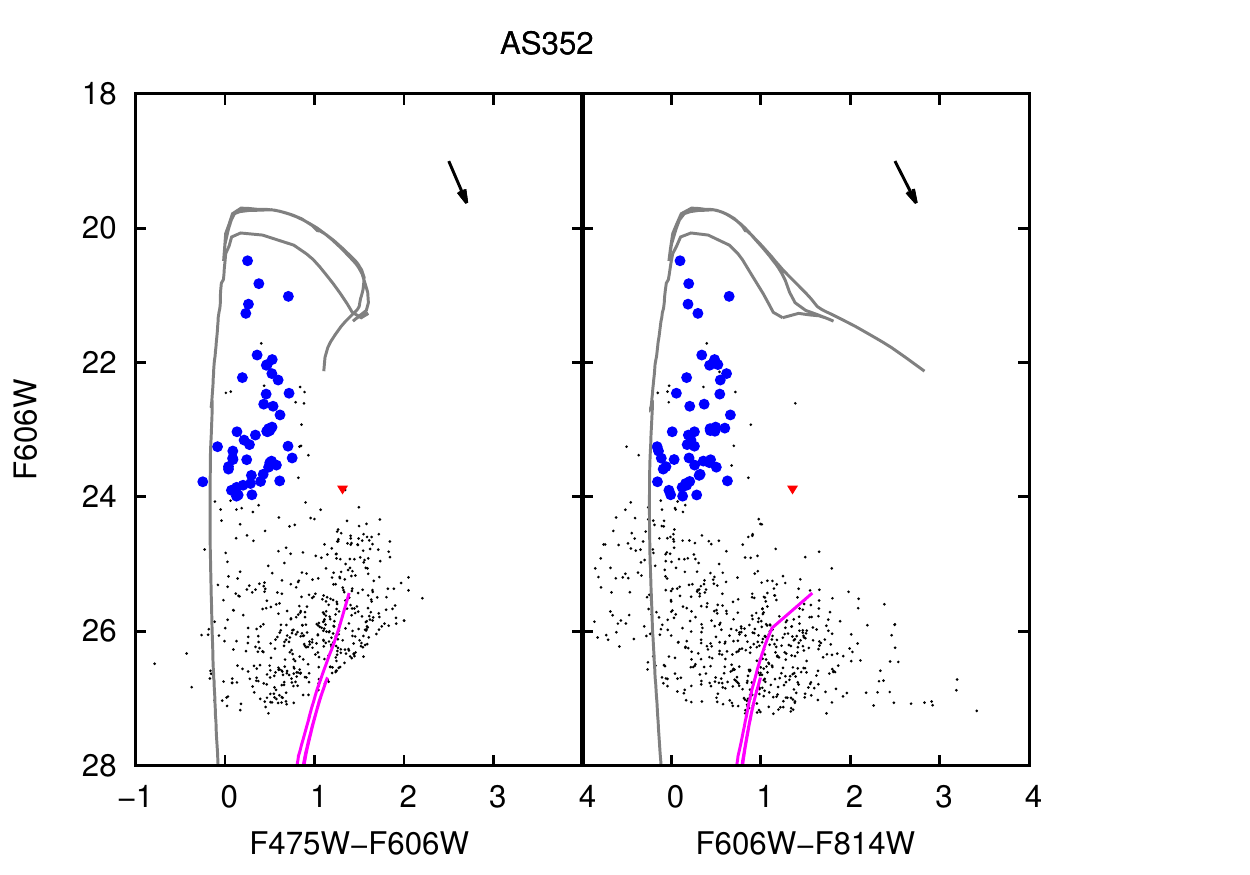}} \\
{\includegraphics[scale=0.65]{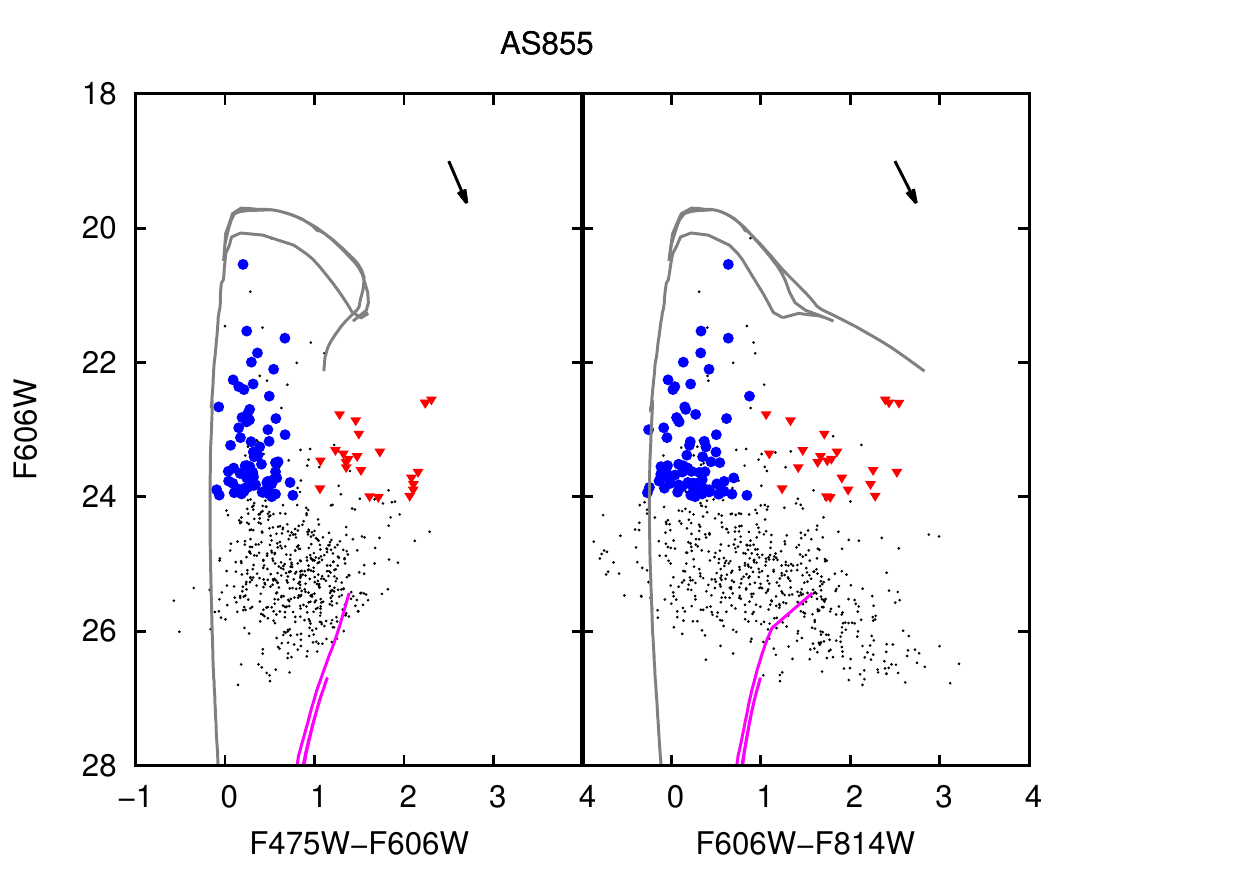}} \\
\caption{Decontaminated CMDs $F606W$ vs. $F475W-F606W$ and $F606W$ vs. $F606W-F814W$ of the star groups AS150, AS352, and AS855, that are located in different galactocentric distances. The meaning of symbols is explained in the text (Sect. \ref{cmd_sect}). The grey and pink lines indicates the isochrones corresponding to $10^7$ yr and $10^9$ yr respectively and metallicity ($Z = 0.0152$). \citep{2014MNRAS.445.4287T}. The arrow indicates the reddening vector. A color version of this figure is available online.}
\label{CMDs}
\end{figure}

We built the decontaminated CMDs, $F606W$ vs. $F475W-F606W$ and $F606W$ vs. $F606W-F814W$ for each star group, taking into account the blue and red population identified in Sect. \ref{blue}.

In Fig. \ref{CMDs} we show the CMDs of three young groups. We overlapped the  evolutionary models of PARSEC version 1.2S corresponding to 10$^{7}$ yr (gray line) and 10$^{9}$ yr (pink line) with solar metallicity for comparison. These models were displacing adopting a distance modulus of $(Vo-M_{V})=27.75$ \citep{2013AJ....146...86T}, a normal reddening law ($R = A_{V}/E(B-V) = 3.1$) and a value for $E(B-V) = 0.016$ corresponding to the foreground reddening toward NGC~253 \citep{2011ApJ...737..103S}. In these diagrams we can note that a small group of \textit{red objects} still prevails. This is because in the right region of the diagrams the stars are more dispersed, causing that the statistical subtraction lose efficiency. Nevertheless, we could see that for the \textit{blue objects} the statistical subtraction is more reliable.

\subsection{Luminosity function}
\label{LFsect}

The $F606W$ LF of the brightest end of each detected young group were build using only the selected \textit{blue~objects} inside the corresponding radius $r$ (see Sect. \ref{blue}) and adopting a bin interval of 0.5 mag.The LF slopes, $\Gamma~=~d~log~N~/~d~F606W$, were determined performing a linear least squared fit. The obtained slopes are listed in Table \ref{tabla1}.

\begin{figure}
{\includegraphics[scale=0.65]{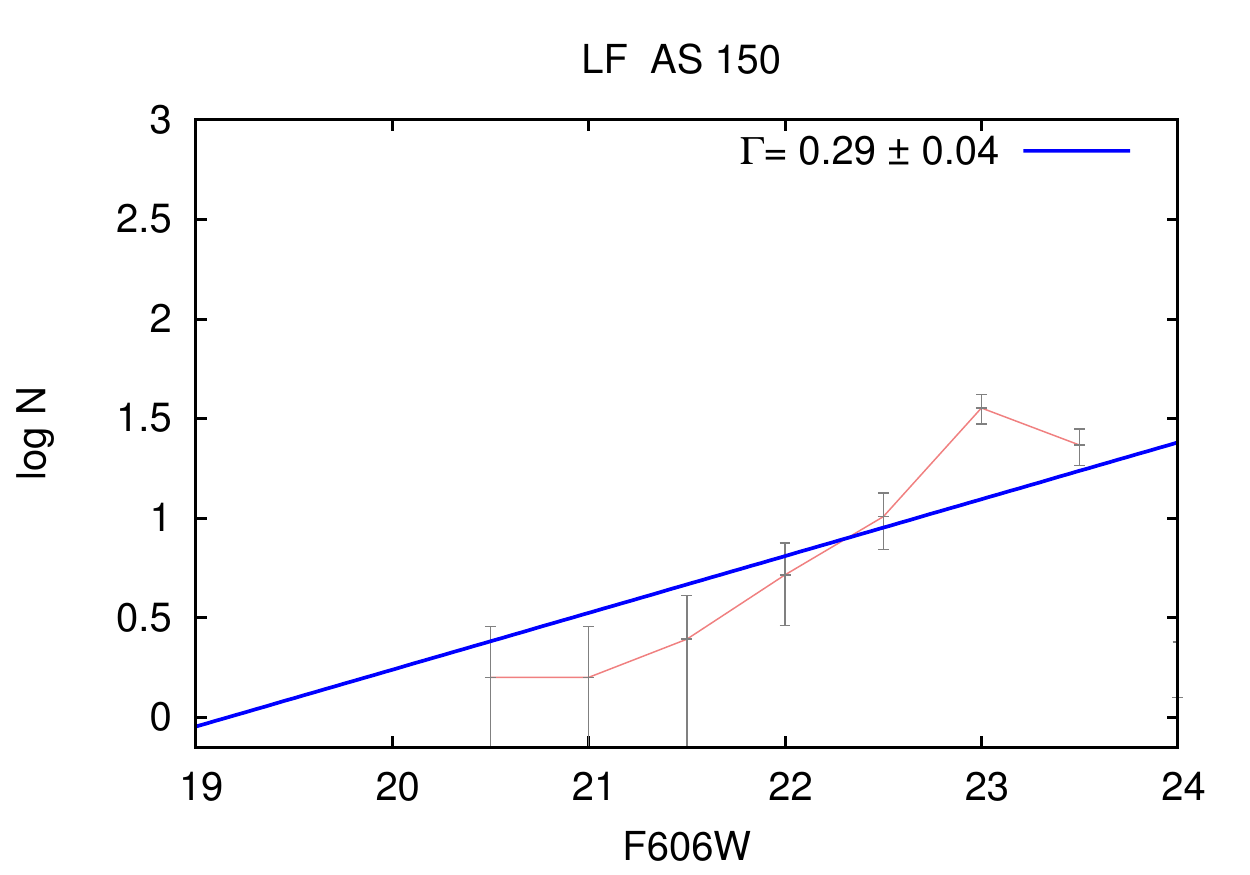}} \\
{\includegraphics[scale=0.65]{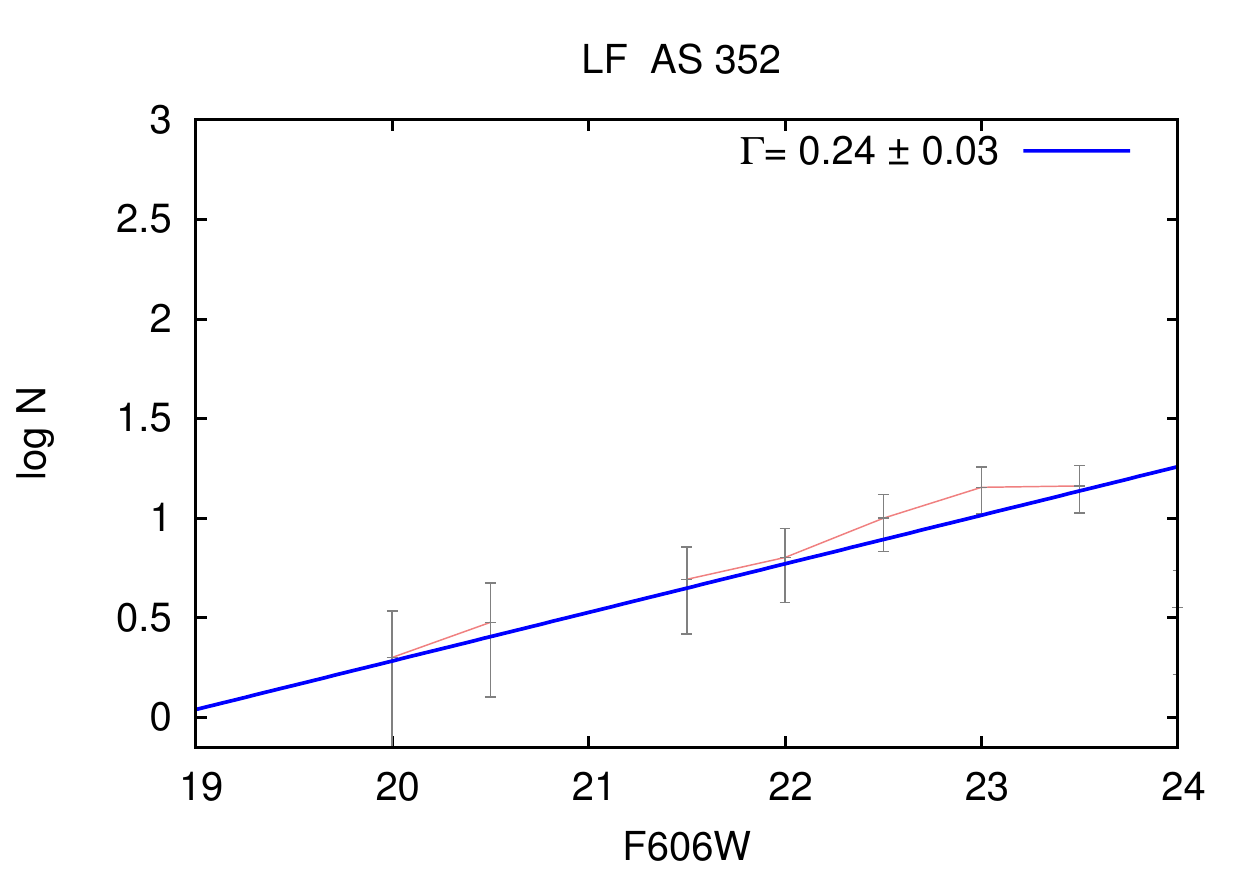}} \\
{\includegraphics[scale=0.65]{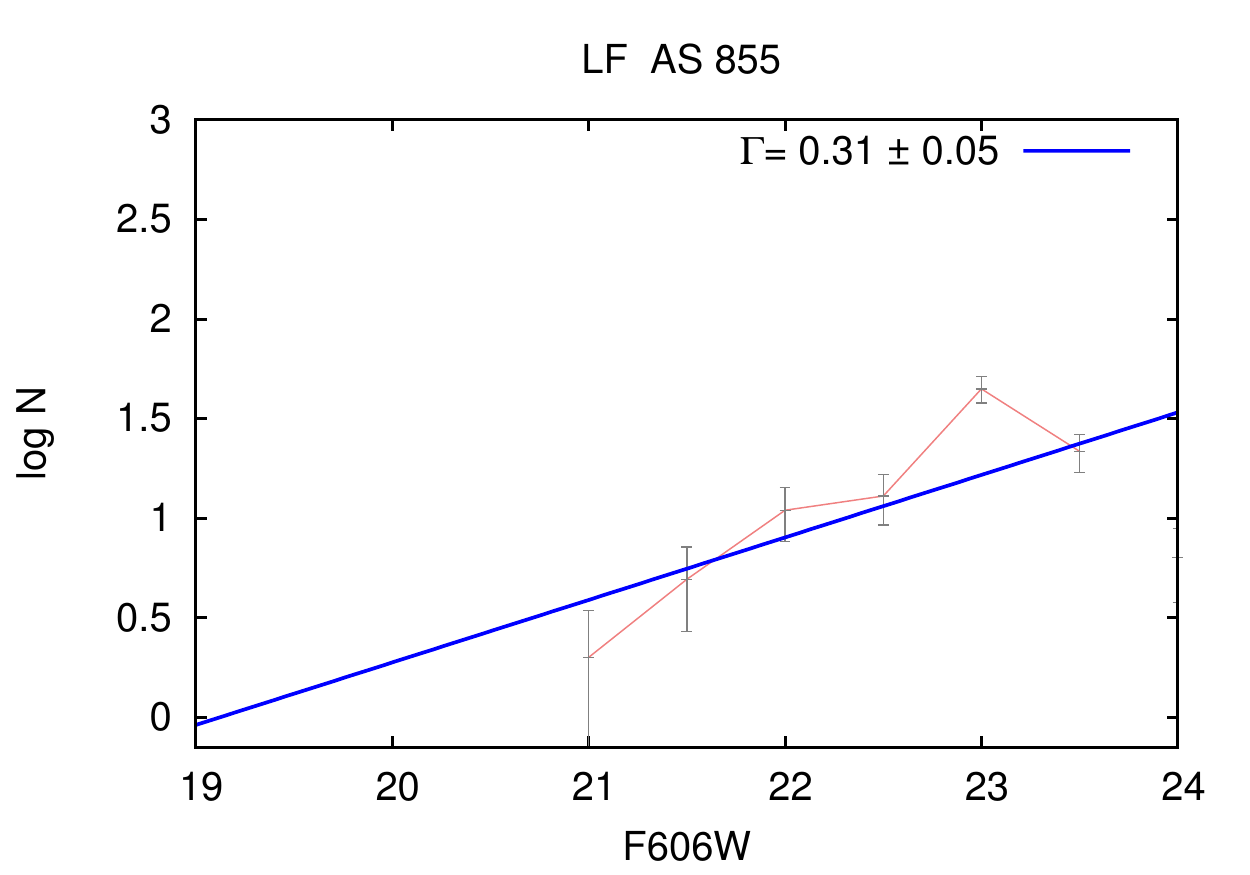}} \\
\caption{LF of three young star groups located at different place in the galaxy, AS~150, AS~352 and AS~855. The line indicates a linear fit over the considered bins (see Sect. \ref{LFsect}), the slope estimated in each case is indicated in the top right corner.}
\label{FL}
\end{figure}

In Fig.~\ref{FL} we present the LFs for the groups AS~150, AS~352 and AS~855, the straight lines indicate the linear fits and the indicated errors correspond to \textit{N$^{1/2}$}, where \textit{N} is the number of stars in each 0.5 magnitude bin. 

The mean slope was derived considering only the groups with more than 30 bright members and err$_{\Gamma}~\le0.05$, obtaining a value of 0.21.

\begin{table*}
\caption{Catalog of young star groups in NGC~253}
\label{tabla1}
\centering 
\begin{tabular}{l c c c c c c c c  c c c r} 
\hline\hline 
name   &   raj2000    &   dej2000  &    r       &    N       &  N$_{dct}$ &  N$_{dct-bri}$ & $F606W_{bb}$ & dens[*/pc$^{3}$] & LF slope ($\Gamma$)  &  err$_{\Gamma}$  &  d$_{GC}$  & $A_v$\\
       &   (deg)      &    (deg)   &   [pc]     &            &            &                &              &                  &                      &                  &     [Kpc]  &       \\   
\hline
 
AS002   &  11.911189  & -25.250842  &   14.80   &    48      &      32      &       18      &     19.99    &      0.00314    &        0.12          &  0.03            &    3.91     &   0.610\\
AS003   &  11.894225  & -25.283620  &   13.77   &    31      &      23      &       19      &     19.85    &      0.00280    &        0.04          &  0.05            &    0.47     &   0.610\\
AS004   &  11.872012  & -25.281292  &   15.15   &    31      &      20      &       17      &     18.45    &      0.00183    &        0.03          &  0.04            &    3.07     &   0.610\\
AS006   &  11.881036  & -25.286077  &   18.59   &    54      &      34      &       12      &     20.77    &      0.00169    &        0.02          &  0.05            &    1.26     &   0.050\\
AS007   &  11.909739  & -25.280824  &   11.36   &    26      &      15      &        9      &     20.94    &      0.00326    &        0.02          &  0.05            &    1.85     &   0.050\\
AS008   &  11.920171  & -25.274954  &    8.26   &    22      &      16      &        9      &     23.12    &      0.00903    &        0.03          &  0.02            &    2.58     &   0.050\\
AS009   &  11.888770  & -25.289815  &   17.90   &    46      &      29      &       22      &     20.26    &      0.00161    &        0.02          &  0.05            &    0.24     &   0.050\\
AS010   &  11.958507  & -25.243423  &   23.41   &   105      &      71      &       44      &     19.18    &      0.00176    &        0.21          &  0.03            &   5.05      &   0.190\\
AS011   &  11.877059  & -25.280397  &   10.33   &    24      &      18      &       11      &     21.80    &      0.00520    &        0.07          &  0.04            &    2.58     &   0.190\\
AS012   &  11.910848  & -25.277194  &   12.74   &    35      &      27      &       18      &     19.77    &      0.00416    &        0.10          &  0.05            &    1.72     &   0.190\\
\hline
\multicolumn{13}{l}{The suffix $bri$ indicates bright stars with $F606W < $ 24.}\\
\multicolumn{13}{l}{The suffix $dct$ indicate stars belonging to the decontaminated region.}\\
\multicolumn{13}{l}{$F606W_{bb}$ is the $F606W$ value of the brightest blue object in the group.}\\
\multicolumn{13}{l}{$d_{GC}$ is the galactocentric distance.} \\
\multicolumn{13}{l}{$A_v$ is the characteristic absorption affecting the group.}\\ 
\multicolumn{13}{l}{Here we present the first ten rows, the complete table is only available in electronic form.}
\end{tabular}
\end{table*}

\section{Discussion}
\label{discussion}

\subsection{Comparison with NGC~300}
\label{crowding}


We compared our detected young groups with those identified in NGC~300, which were studied  by \cite{2016A&A...594A..34R} using similar observational data that in the present work. 

\begin{figure}
{\includegraphics[scale=0.65]{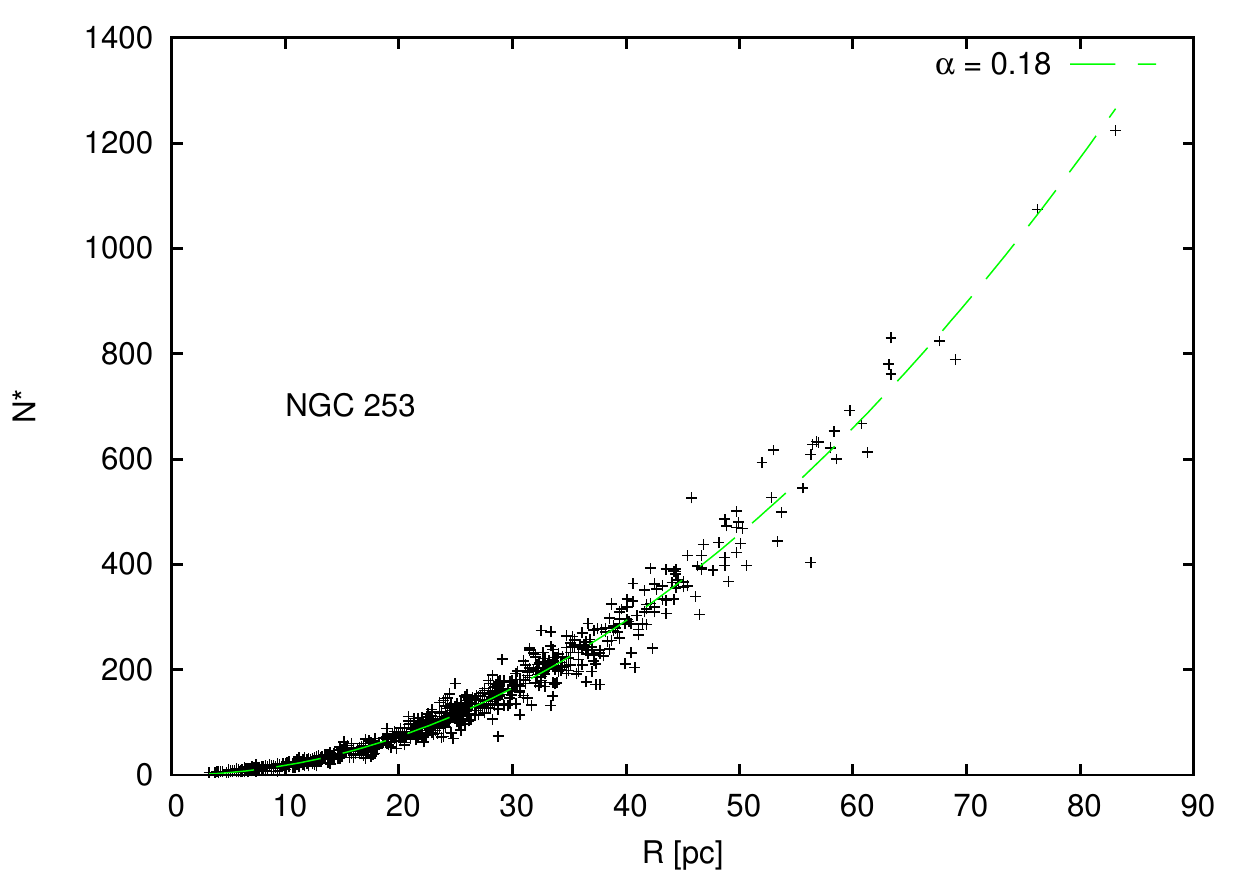}} \\
{\includegraphics[scale=0.65]{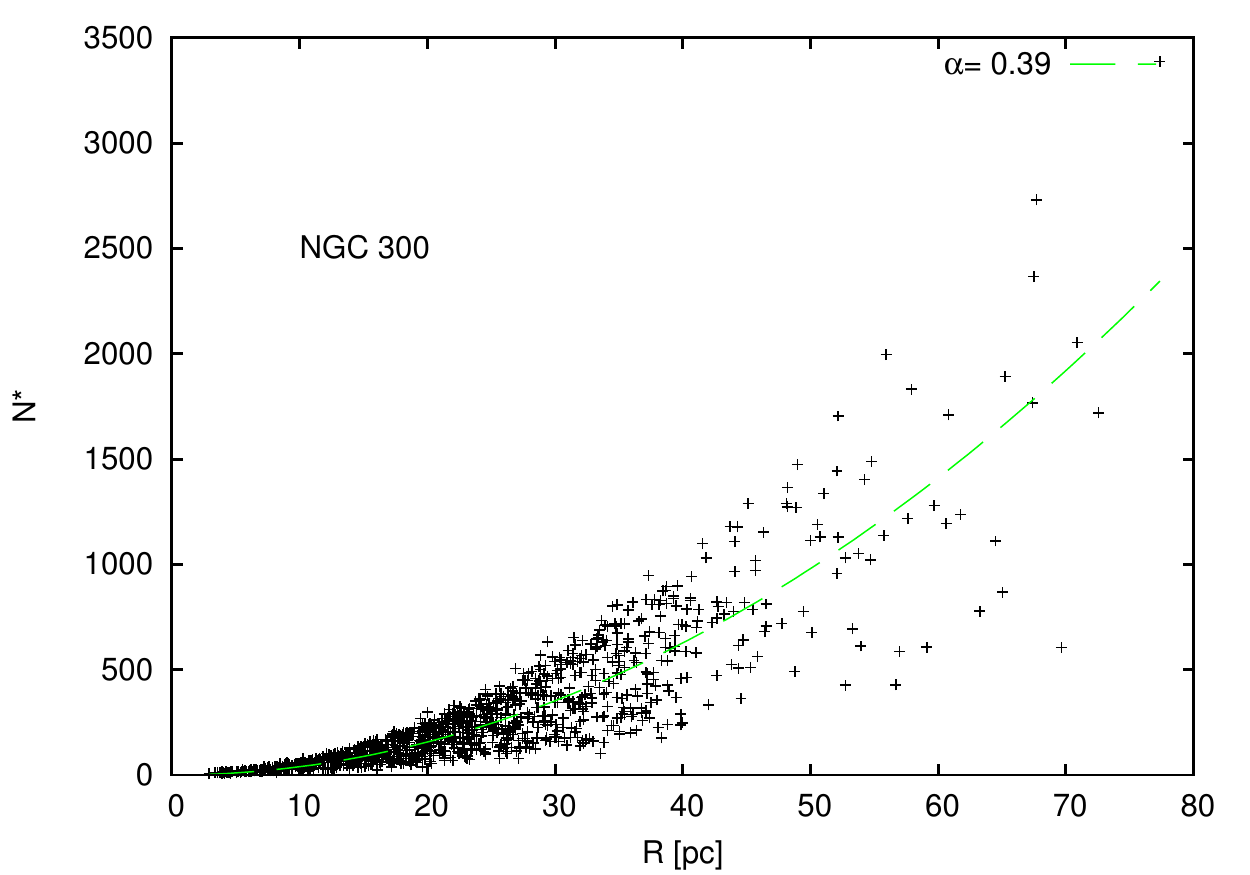}}
\caption{Number of stars members vs. radius of young groups, for NGC~253 (top) and NGC~300 (bottom). The dashed line correspond to a quadratic fit over the data.}
\label{radio_n}
\end{figure}

In Fig.~\ref{radio_n} we plotted the number of stars members versus the size of the groups in each galaxy. To be consistent with \citet{2016A&A...594A..34R} we adopted in this figure a radius of 1$\sigma$ (the radial standard deviation of the groups member location). We performed a quadratic fit to the data, in order to compare the numbers of members detected in groups of the same size in both galaxies. The least squared fit was performed taking into account the groups with sizes larger than 9~pc, to avoid the most crowded regions. We found a ratio of $\sim$ 2 between the quadratic coefficients ($\alpha$) of each galaxy. In other words, the associations in NGC~300 would appear to contain twice as many stars as NGC~253.

However, NGC~300 is located almost half the distance to NGC~253, therefore in this galaxy we were reaching less luminous objects than in NGC~253, or less massive main sequence stars. Thus, it was expected that we could detect more members in the NGC~300 associations, but this was only a distance effect. 


\subsection{Hierarchical structure in NGC~253}
\label{hierarchical}

\begin{figure*}
\centering
\includegraphics[width=18cm]{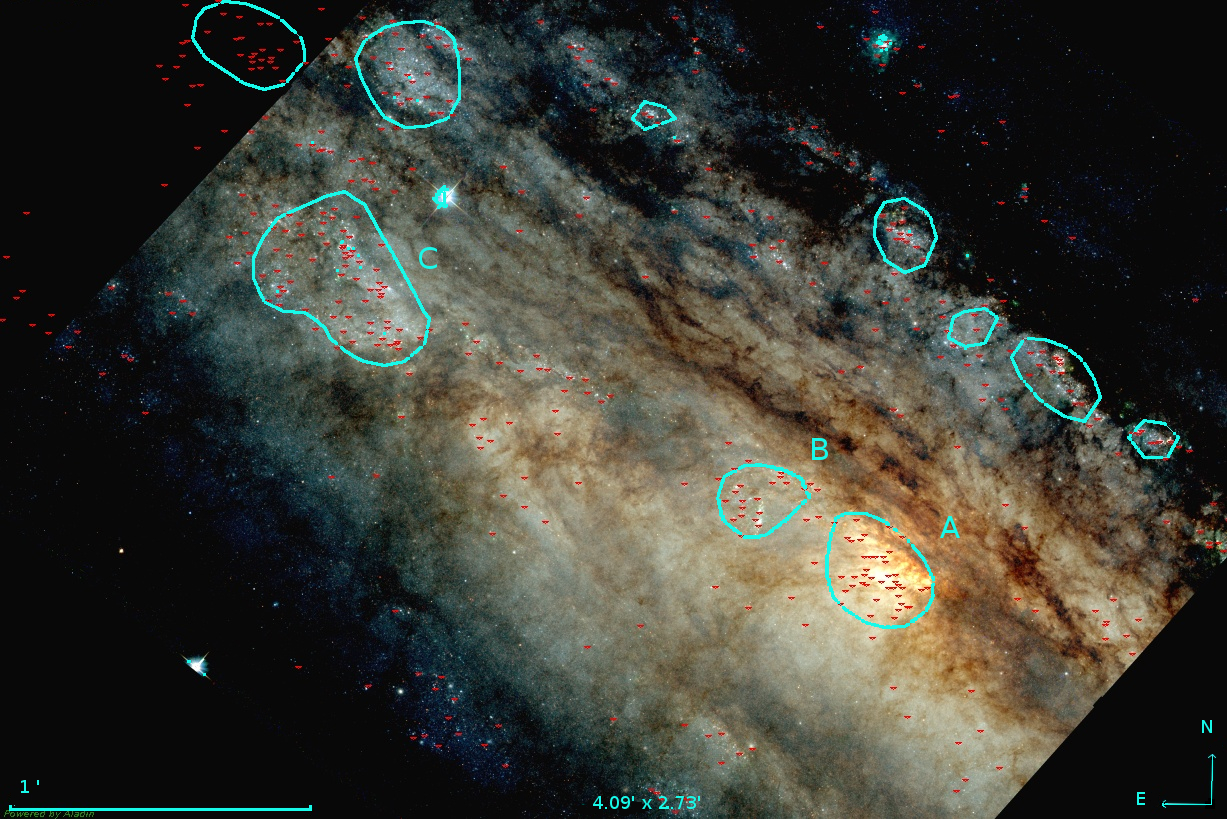}
\caption{ACS/WFC color image of the field 5, obtained from the combination of the filter F475W in blue, F606W in green and F814W in red using ALADIN. The inverted triangles indicate the center of the individual groups detected by the PLC method, the contours are the densest that appear in Figure 8 Fig.~\ref{Dmap} and indicate stellar complexes (see Sect.~\ref{hierarchical} and \ref{distribution}). A color version of this figure is available online.}
\label{field5}
\end{figure*}

As was mentioned in Sect.~\ref{intro}, star forming regions exist in a wide range of sizes, from the compact star cluster to the spiral arms of a galaxy. Using the PLC and the stellar density map (Sects.~\ref{secPLC} and \ref{densitymap}), we have searched and identified an important number of young star groups whose sizes and amount of members revealed they include different kind of structures, from simple open clusters to important complexes of several OB associations. In Fig.~\ref{Dmap} we could see how the largest and most scattered young structures enclose the densest and compact ones, going through different density levels, indicated by the different contours. In fact, if we plot the PLC groups together with the contours obtained from the density map, we see a large number of these groups within the densest contours (see Fig.~\ref{field5}), suggesting that they enclose stellar complexes. 

\begin{figure*}
\centering
\includegraphics[width=18cm]{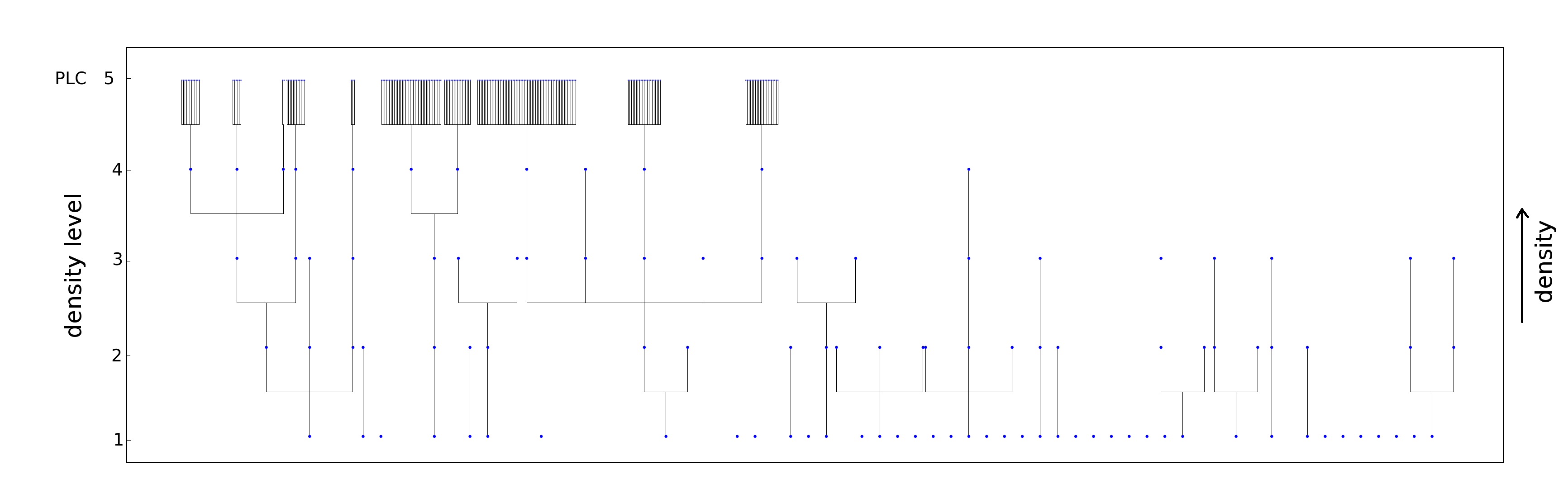}
\caption{Dendrogram of the young stellar structures detected at different density levels. Levels 1~-~4 correspond to 40, 80, 110, and 145 stars per bin of $ 8~\times~8~$arcsec$^2$, respectively. Level 5 corresponds to the most compact groups detected using PLC. The small circles indicate different stellar structures. Structures which are related in a hierarchical way are connected by solid lines.}
\label{dendrogram}
\end{figure*}

This behavior of the young population is known as hierarchical structure \citep[e.g.][]{2000prpl.conf..179E}. We can see the relationships among these structures in Fig.~\ref{dendrogram}, in which we presented the corresponding tree diagram or dendrogram. In the bottom, the diagram starts with the structures detected in the lower density level (40 stars per bin of $ 8~\times~8~$arcsec$^2$, black contour in Fig.~\ref{Dmap}). Most of these structures divide themselves in denser and smaller systems that are detected in the second density level (80 stars per bin of $ 8~\times~8~$arcsec$^2$, blue contour in Fig.~\ref{Dmap}). This behavior is repeated it turns through the more dense levels (110 and 145 stars per bin of $8~\times~8~$arcsec$^2$, red and turquoise contour respectively). Finally, we add a fifth level which correspond to the groups detected with the PLC method and contained in the structures detected in level 4. It is not possible to detect such groups in the density map because the map pixels sizes is 2 arcsec which is comparable to the groups size. 
 
\subsection{Distribution of the young star groups in the galaxy}
\label{distribution}

As was indicated in Sect.~\ref{secPLC}, in Fig.~\ref{asoc} we show the distribution of the identified groups on an infrared WISE image of the galaxy. Therefore, it is possible to note that the stellar groups are mainly located over some special features of NGC~253, they are I) the nuclear region, II) the observed extreme of the bar, III) a ring like structure enclosing the bar, and IV) the spiral arms. All these structures are prominent in the near infrared spectral range \citep[see Fig.~4 of][]{2014A&A...567A..86I}.

In particular, we found two stellar complexes with almost twenty members (individual young groups) near the central region of the galaxy (contours A and B in Fig.~\ref{field5}). The closer to the center (complex A) is within the 300~pc of starburst activity, and it is coincident with the super star cluster described in several works \citep[eg.][]{2009ApJ...697.1180K,2016ApJ...818..142D}. Through the PLC method we could identify several individual star groups in this region, suggesting that it is in fact a stellar complex rather than a super cluster. The same conclusion was arrived by \citet{2016ApJ...818..142D} who found star formation in different locations inside this region.  
  
We also can see in Fig.~\ref{asoc} a large stellar complex containing several groups at the edge of the bar, this complex is indicated with the contour C in Fig.~\ref{field5},  in this last optical image the bar is totally obscured by the dust and only are visible the star forming region in the bar edge. This knot and the other extreme of the bar were identified by \citet{2014A&A...567A..86I} as bright points, in near infrared images of VISTA\footnote{http://www.vista.ac.uk/}, that connect the bar with the outer spiral arms. Their analysis suggest these knots as regions of local star formation instead of being the typically ansae observed in other barred galaxies, which is consistent with ours detections.

On the other hand, there was expected to find young groups in the ring structure, since this region is associated with active star formation as was suggested by the $H_{\alpha}$ maps of the region \citep{1996AJ....112.1429H}. The origin of this structure, with a radius between 2.6 and 3.1 kpc, is not yet well understood. According to \citet{2014A&A...567A..86I} it is probably the result of a merge with a small satellite or, alternatively, it is an intermediate phase in the bar formation.

\subsection{General properties of the young star groups}

\begin{figure}
\resizebox{\hsize}{!}{\includegraphics{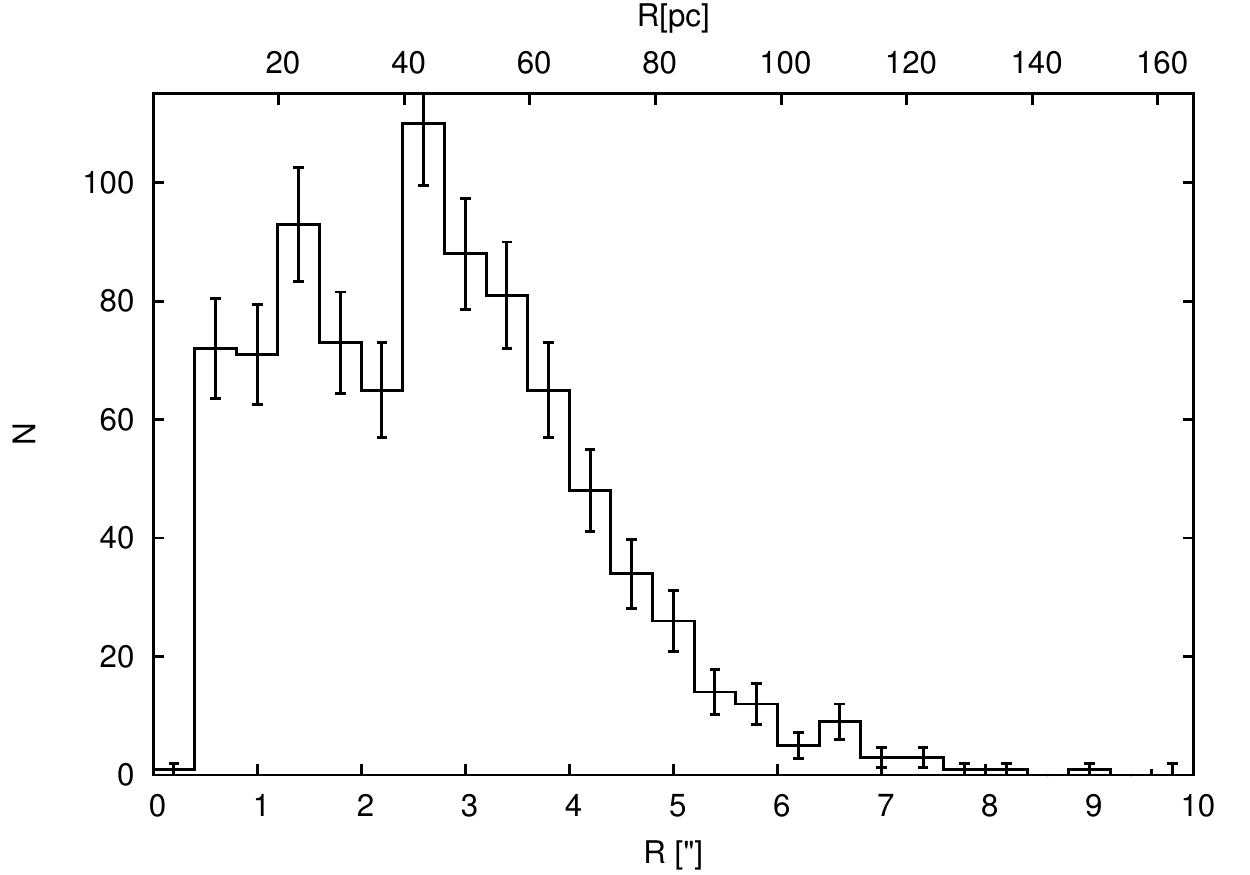}} 
\caption{Size distribution of young star groups in NGC~253. Sizes in parsecs (upper edge) was computed adopted a distance of 3.56~Mpc \citep{2013AJ....146...86T}.}
\label{H_radio}
\end{figure}

In Fig.~\ref{H_radio} we show the size distribution of the star groups, we can see that the radius ranges from 5 to 150~pc with a peak close to 40 pc. We found an average radius of $\sim$ 47~pc. These values are in good agreement with the OB association size found in others galaxies, using automatic search methods and HST images. \citet{2016A&A...594A..34R} found a value of 25~pc for the mode and the average radius of young stars groups in NGC~300, although these values are lower than ours, we must consider that they used values of radius equal to 1$\sigma$ instead of 2$\sigma$ as in this work. \citet{1998AJ....116..119B} studied the OB associations in seven spiral galaxies, NGC~925, NGC~2090, NGC~2541, NGC~3351, NGC~3621, NGC~4548, and M101, finding size distribution peaks between 25-45~pc and average radius a between 25-60~pc. In our Galaxy  these values are somewhat smaller. \citet{1995AstL...21...10M} found a maximum in the size distribution of 15 pc and an average radius of 20 pc in the OB associations with distances within 3 kpc of the Sun. For example, the associations in the Orion region Ori OB1a and Ori OB1b have radius of 37 pc and 17 pc respectively \citep{2008hsf1.book..838B}, the associations in Cygnus region have an extended range of radius from 10 to 70 pc \citet{1992A&AS...94..211G}.

\begin{figure}
{\includegraphics[scale=0.50]{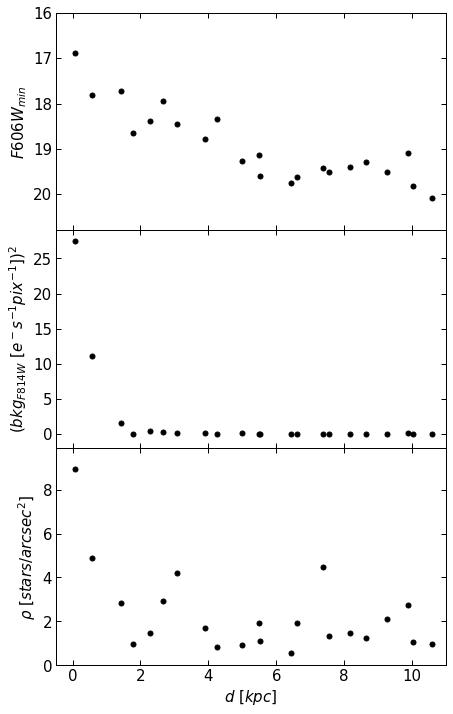}} \\
\caption{({\it Top panel:}) Magnitude of the brightest star in the stellar groups, ({\it Middle panel:}) the red ($F814W$) background level and ({\it Bottom panel:}) the stellar density of stars brightest than $F606W=24$, with respect to galactocentric distance. In the three panels was adopted a bin size of 0.5~kpc and were considered the same stellar groups.}
\label{triple}
\end{figure}

In Fig.~\ref{triple} we presents the trend of several parameters against the galactocentric distance. In the upper panel we show the $F606W$ magnitude value of the brightest star in the stellar groups for each $0.5~$kpc bin; the middle panel presents the behavior of the red ($F814W$) background level measured on the  ACS/WFC drizzle images; and the bottom panel indicates the stellar density values of bright stars. Both, the background levels and the stellar density values correspond to the same stellar groups indicated in the upper panel.

These figures allowed us to notice that the brightest stars in the stellar groups was decreasing up to distances $\sim 6~$kpc and then remained approximately constant. The behavior at the central part of the galaxy could be due to a blending/crowding effect, a change in the stellar density since it directly affect the chance to get a bright star for a given initial mass distribution, or a flatter slope of the mass distribution itself. The middle panel of Fig.~\ref{triple} was adopted as an indicator of the blending behavior (\citealt{1998AJ....115.2459R}) and revealed that this problem is not important in most regions of the galaxy, as was indicated in Sect.~\ref{blending}, but it could be important for galactocentric distances lower than $2~$Kpc. On the other hand, the bottom panel of Fig.~\ref{triple} revealed a changing behavior of the stellar density. These results suggested that the trend observed in the upper panel could be mainly due to changes in the stellar density. However, in the central part of the galaxy, the blending effect and the high stellar densities would be a problem.



\begin{figure*}
{\includegraphics[scale=0.6]{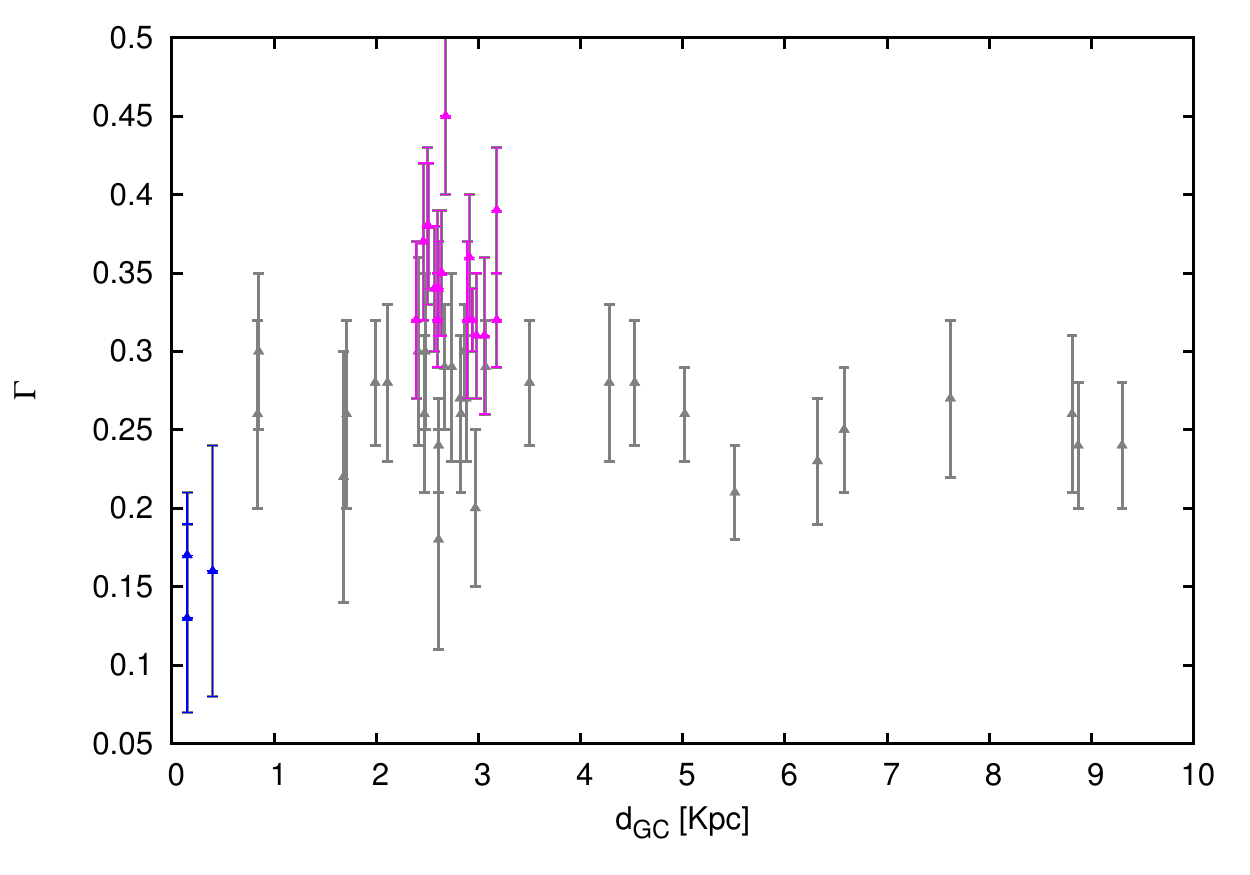}}
{\includegraphics[scale=0.25]{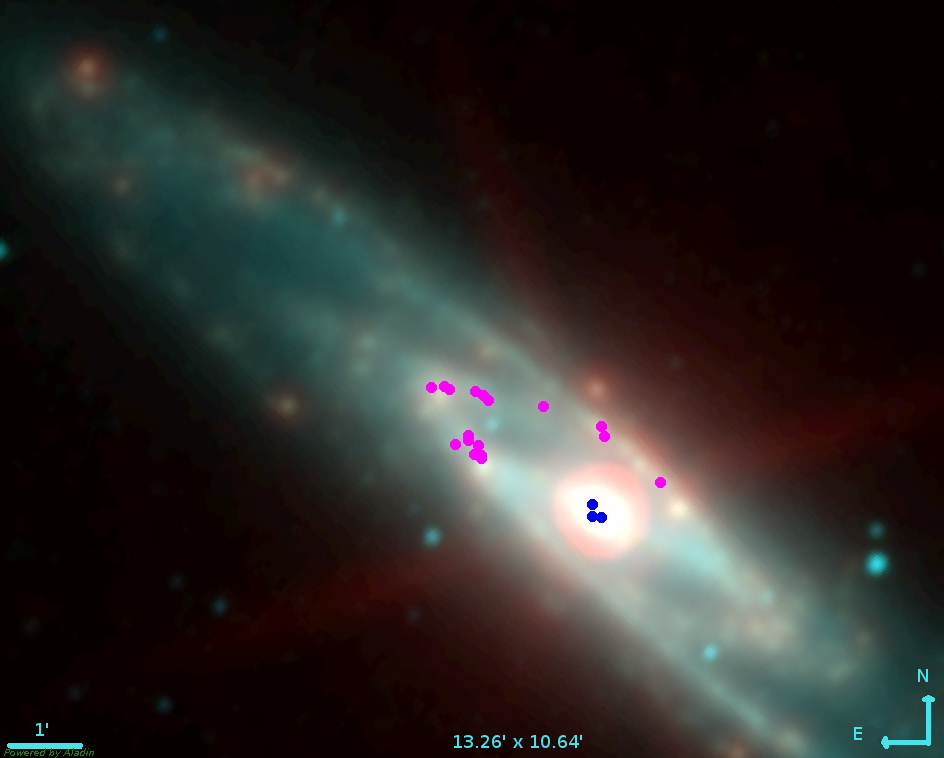}}
\caption{{\it Left:} Behavior of the LF slope with the galactocentric distance. Pink dots indicate groups with $\Gamma$ values greater than 0.3. Blue points are groups with $\Gamma$ values lower than 0.18. {\it Right:} Location of pink and blue dots over a Wise IR image of NGC~253. It can be clearly seen that the associations with low $\Gamma$ values are found in the galactic core, while the ones with high values are located over an annular structure surrounding the galactic core.}
\label{FL_dist}
\end{figure*}

Regarding the stellar mass and bright distribution of the stellar groups, \citet{1998AJ....116..119B} estimated an average $V$ LF slope of 0.61 for the spiral galaxies mentioned at the beginning of this section. This value is considerably higher that our computation of 0.21. This discrepancy may be partly due to the different magnitude cutoff adopted in the slope measure \citep[$M_{V}\le$-4.76 in ][and $M_{F606W}\le$-3.81 in our work]{1998AJ....116..119B}. Additionally, the \citet{1998AJ....116..119B} galaxies are much more distant than NGC~253 (between 6.7 and 14.5~Mpc), so it is probably that a blending effect is present affecting the objects magnitudes and consequently the LF shape. Studying only the most numerous groups with low errors in the fitted slope, that is those with more than 50 bright blue stars resulting from statistical decontamination and errors in the slopes lower than 0.09, we found that the groups with low slope values were located near the galactic center. On the other hand, groups with the highest slope values were located between $\sim$ 2-3.5 Kpc. The left panel of Fig.~\ref{FL_dist} shows the relationship of $\Gamma$ and the galactocentric distance, in blue color were marked the groups with slope values lower than 0.18 and the slopes greater than 0.3 were marked in pink color. In the right panel of Fig.~\ref{FL_dist} we overlapped these associations on a IR WISE image of the galaxy. In this figure we can note that the groups with low values of the slope are near the galactic center, within the starburst region and inside complex A (see Fig.~\ref{field5}). These low values indicate that the LF slope is flatter than in the groups of other regions, so there would be more bright stars than in other groups, however this region is well correlated with the behavior presented in the middle panel of Fig.~\ref{triple}, therefore these values could be partially affected by the presence of blending. On the other hand, all the groups with slopes greater than 0.3 are over the annular structure that surrounds the bar and the nucleus, the LFs of these associations are characterized to have a large number of stars of weaker magnitudes.


Additionally, We estimated a mean density of 0.0006 stars/pc$^{3}$ taking into account groups with more than 30 bright stars. It should be noted that this value was estimated counting stars with M$_{F606W} \le -0.8$, which approximately correspond to the spectral type \textit{B6}.

\section{Conclusions}
\label{conclusions}

Using  ACS/HST images we searched and identified young star groups in the starburst galaxy NGC~253. For this task, we first derived the absorption affecting different regions of the galaxy. After correcting by this effect, we used the PLC method over the blue bright objects to identify young groups and build the density map to detect larger young structures. A special designed code was run over the detected groups to estimate their fundamental parameters and to build their corresponding CMDs and LFs. 

We constructed a catalog containing the characteristics of the 875 detected young groups. This catalog presents coordinates, sizes, number of members, densities, LF slopes and galactocentric distances.


Our study revealed that the groups detected are located over prominent structures of the galaxy, its nuclear region, the edge of the bar, a ring like structure that enclose the bar, and the spiral arms, confirming that all of them are star forming regions. The nuclear region of this galaxy had been greatly studied due to its starburst activity, it had been associated to a super star cluster by several authors. Nevertheless, we identified almost 20 different groups in this regions, suggesting that it is in fact a stellar complex. Additionally we found that this young population have a hierarchical behavior, in which the smaller groups are contained in larger structures. 

The groups size distribution revealed a peak near 40~pc and an average radius of 47~pc, this values are consistent with those found in other galaxies. We also estimate a mean value of the $F606W$ LF slope of 0.21, with lower slope values in the galactic center and higher values in the ring structure, and an average density of 0.0006~stars/pc$^{3}$ for stars considered earlier than B6.

\section*{Acknowledgements}
We thank the referee for helpful comments and constructive suggestions that helped to improve this paper.
MJR and GB acknowledge support from CONICET (PIP 112-201101-00301). MJR is a fellow of CONICET. This work was based on observations made with the NASA/ESA Hubble Space Telescope, and obtained from the Hubble Legacy Archive, which is a collaboration between the Space Telescope Science Institute (STScI/NASA), the Space Telescope European Coordinating Facility (ST-ECF/ESA) and the Canadian Astronomy Data Centre (CADC/NRC/CSA). Some of the data presented in this paper were obtained from the Mikulski Archive for Space Telescopes (MAST). STScI is operated by the Association of Universities for Research in Astronomy, Inc., under NASA contract NAS5-26555. Support for MAST for non-HST data is provided by the NASA Office of Space Science via grant NNX09AF08G and by other grants and contracts. This publication makes use of data products from the Wide-field Infrared Survey Explorer, which is a joint project of the University of California, Los Angeles, and the Jet Propulsion Laboratory/California Institute of Technology, funded by the National Aeronautics and Space Administration. This research has made use of "Aladin sky atlas" developed at CDS, Strasbourg Observatory, France.

\bibliographystyle{mnras}
\bibliography{references} 

\bsp	
\label{lastpage}
\end{document}